# Numerical modelling and simulation on nanosecond laser-target interactions

**Jian Wu, Ying Zhou, Minxin Chen, Xingwen Li***

[1] State Key Laboratory of Electrical Insulation and Power Equipment, Xi'an Jiaotong University, Xi'an 710049, China

*xwli@mail.xjtu.edu.cn

## Abstract

Nanosecond lasers are widely used in industrial applications as they are relatively inexpensive, and their compactness and robustness are an advantage. Much experimental work has been carried out to understand deeper the interaction between the nanosecond laser pulses and the targets, as these are complex, transient processes with spatial inhomogeneities. Beside the experiments, the modeling and numerical simulation on the laser interaction with the target are also crucial for understanding the dynamics of laser-material interactions and for optimizing laser processing applications. In this review, the progress of numerical modelling and simulation on nanosecond laser-target interactions are summarized from the aspects of laser-target interactions and target-plasma interface, laser-plasma interactions and plasma radiation, and numerical models on different scales with artificial intelligence advancing. The laser ablation, mass and energy transfer, and mechanical coupling are discussed in the aspect of the nanosecond laser-target interactions and target-plasma interface. The plasma expansion, plasma ionization and recombination, and plasma radiation are discussed in the aspect of the nanosecond laser-plasma interactions and plasma radiation. Then the numerical advances, including microscopic approaches based on molecular dynamics, mesoscopic approaches based on kinetic and statistical physics, macroscopic approaches based on fluid dynamics, and numerical simulations with machine learning are discussed. Finally, the challenges currently being encountered by numerical modelling and simulation on nanosecond laser-target interactions and its potential development direction are considered.

**Abbreviation**

| Abbreviation | Text | Abbreviation | Text |
|---|---|---|---|
| CR | Collisional-radiative | LIBS | Laser-induced breakdown spectroscopy |
| CTE | Complete Thermodynamic Equilibrium | LIP | Laser-induced plasma |
| DSMC | Direct Simulation Monte Carlo | LSP | Laser shot peening |
| EEDF | Electron energy distribution function | LTE | Local thermodynamic equilibrium |
| EOS | Equation of state | MC | Monte Carlo |
| FDM | Finite difference method | MD | Molecular dynamics |
| FEM | Finite element method | MHD | Magnetohydrodynamics |
| FVM | Finite volume method | PDEs | Partial differential equations |
| HD | Hydrodynamics | PIC | Particle-in-Cell |
| IR/Vis/UV | Infrared/visible/ultraviolet | PINNs | Physics-informed Neural Networks |
| LBM | Lattice Boltzmann method | PLD | Pulsed laser deposition |

## 1. Introduction

The laser-target interaction can produce a predefined region with wide range of parameters in temperature, density, pressure, charge states based on the irradiation conditions. It has been receiving wide interest for both cutting-edge scientific researches and practical applications such as high-energy density physics (HEDP) [1, 2], advance light sources [3, 4], laser processing [5], surface analysis [6] and etc. It is a transient, inhomogeneous and complex process and greatly influenced by multiple factors, such as the laser parameters (fluence, beam dimensions, duration, time-shape), target characteristics (physical, chemical, mechanical parameters), and the surrounding medium. The physics behind the interaction revolving around the multifaceted aspects of photon-matter coupling at high photon flux with varying laser intensity and pulse widths. For the nanosecond laser pulse, the absorption process is mainly induced by thermal ionization, and then ablation and plasma formation due to the surface heating. The ablation on the target surface is primarily a non-thermal process when the laser pulse width is reduced to the picosecond or femtosecond scale. The non-thermal processes mainly include multi-photon absorption and ionization [7, 8], tunneling [9], Coulomb explosion [10], and electron dynamics [11].

Due to its advantages of relatively inexpensive, compact and robust, the nanosecond lasers are more widely used in industrial applications. The laser energy deposited in the target is used in the material processing specifically for cutting, drilling, welding, marking, and engraving in the field of automotive and aerospace, electronic, and etc [12]. The plasma plume formed are used for various analytical tools such as laser-induced breakdown spectroscopy (LIBS), laser ablation inductively coupled plasma mass spectrometry (LA-ICP-MS), laser ablation laser-induced fluorescence (LA-LIF). The particles produced during condensation are used to produce nanoparticles, the deposition of layers and coatings [13].

The interaction of nanosecond laser with the solid target is mainly caused by thermal effects. As the laser intensity increases, the interaction between the laser and solid target transforms from target heating and thermal-elastic deforming, to target melting, and then to target ablation followed by plasma expansion. While in most applications, the laser in low to moderate intensity ranges ($10^8$-$10^{15}$ W/cm$^2$) but above the melting or even plasma breakdown threshold of materials has been widely used. Under these conditions, laser energy is transformed into electronic excitation energy and then transferred to lattices of materials through electron lattice interaction when a nanosecond laser irradiates matter. The laser energy deposition on the target surface can lead to the heating, melting, vaporization, phase explosion and ionization of the material. Then vapor/plasma forms, expands, and interacts with the laser and the ambient gas. Finally, the phenomena such as particle formation and surface modification occur with the emission of plasma radiation [14]. Meanwhile, the mechanical effects, such as material expansion, thermoelastic and plastic stresses may occur in the material [15]. It should be noted that, in this review, ablation processes dominated by photothermal mechanism with plasma generation are mainly discussed. It involves a solid-phase heating, melting-phase generating, and a thermal influence on the adjacent material region. The development of the materials removed from the surface is then investigated in a gas/plasma phase. This photothermal mechanism differs from the photochemical mechanism, in which mainly molecules are fragmented by breaking bonds without significant heating (e.g. ultraviolet short-pulse lasers in eye surgery, chemical decomposition of organic substances). [16, 17].

Due to a complex process with its transient nature combined with spatial inhomogeneities, a lot of work have been carried for a deeper understanding the interaction of the nanosecond laser and the target in experiments. Various plasma diagnostics are applied to measure the plasma parameters and reveal the underlying physical processes. Due to their small size and transient and inhomogeneous nature, the plasma diagnostic tools used in the experiments needs to have high spatial and temporal resolutions. Harilal reviewed the optical diagnostics of laser-induced plasma (LIP), such as optical spectroscopy (emission, absorption, and fluorescence), as well as passive and active imaging and optical probing methods (shadowgraphy, Schlieren imaging, interferometry, Thomson scattering, deflectometry, and velocimetry) [18]. There are also many works that focus on tuning the interaction to overcome the limitations due to signal uncertainty, and matrix effects. [19].

Beside the experiments, the modeling and numerical simulation on the laser interaction with the target are also crucial for understanding the dynamics of laser-material interactions and for optimizing laser processing applications. A growing interest in modeling arising from applications including prediction of crater morphology in laser processing, formation and motion of multiple ion species for ion sources, or analysis of plasma radiation. Normally, the thermal model is used for the ablation, and radiative-hydrodynamical models are used to calculation the expansion and radiation of the plume [20, 21]. Recently, more accurate dynamic equations of state and transport parameters are adopted to the calculation of ablation craters. Non-stoichiometric ablation is used to calculate the mass and energy transfer from target to plasma. Collisional radiative models and multi-fractal fluid models are applied to describe the non-equilibrium and multi-species processes. Meanwhile, the lattice Boltzmann method (LBM), Monte Carlo (MC), and advancing numerical simulations with machine learning are also applied.

In this review, we focus on the numerical modeling and simulation of nanosecond laser-target interactions with laser intensities in the low to medium range, which are mainly used in industrial applications. The organization of this review is as follows. Section 2 provides the progresses on the modelling and simulation of nanosecond laser-target interactions and target-plasma interface, including laser ablation, mass and energy transfer, and mechanical coupling. In Section 3, the advances on the nanosecond laser-plasma interactions and plasma radiation are presented, including the plasma expansion, plasma ionization and recombination, and plasma irradiation. Section 4 discusses numerical advances on different scales, including microscopic approaches based on molecular dynamics (MD), mesoscopic approaches based on kinetic and statistical physics, macroscopic approaches based on fluid dynamics, and advancing numerical simulations with machine learning. This review ends with a summary (Section 5) on the progress of numerical modelling on nanosecond laser-target interactions and challenges. Detailed experimental configurations in most of the literature are added in parentheses to the end of the statement in the order of laser, target, and environment. Some of the numerical methods are also expanded for a better understanding of the modeling process. The order of additions is specifically defined as (model; laser: wavelength, pulse width, energy (or intensity), spot size (or diameter), repetition frequency; target; environment). Options of which are skipped if they are not explicitly stated in the literature.

## 2. Nanosecond laser-target interactions and target-plasma interface

The interactions between nanosecond laser pulses and targets during laser ablation are accompanied by phase changes such as melting, vaporization, phase explosion, and appearance of thermomechanical stresses, etc. (see Figure 1) These processes determine the ablation threshold and surface morphology of the target. In Section 2.1, models describing laser ablation at different laser irradiance inputs are presented, with inversion and inverse measurements of material physical properties to some extent through thermal coupling also noted. In Section 2.2, the mass and energy transfer of vapor components ablated into the environment in relation to gas-solid-liquid surface tracking is also further discussed, and more specifically, non-stoichiometric ablation and evaporation time discrepancies are also included for multi-component materials.

In Section 2.3, it is also concerned with the thermo-mechanical coupling that occurs during nanosecond laser ablation, which has been measured and repetitively confirmed in a number of relevant experimental situations, including thermo-mechanical stresses and ultrasonic waves within the target, as well as the recoil pressure and shockwave expansion phenomena during plasma formation. These mechanical effects assist in explaining the mechanisms of laser cleaning, mechanical damage and nano/microscale reliefs formation, deepening the understanding of the multiple mechanisms that arise under nanosecond laser-target interactions.

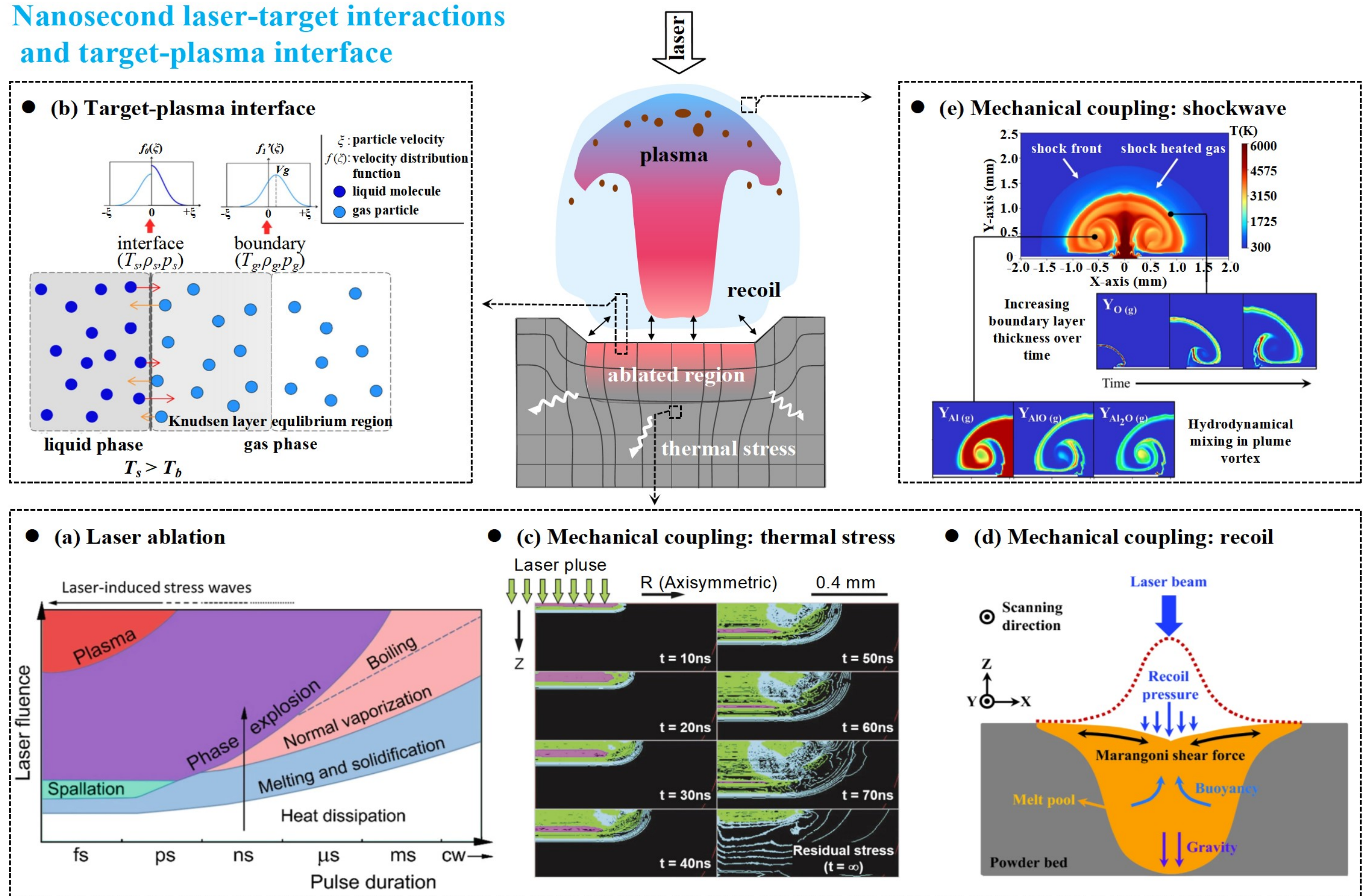

**Figure 1.** Schematic diagram of nanosecond laser-target interactions and target-plasma interface. (a) Laser ablation: illustrating manifestation of thermal processes at laser fluences typical for material processing. Reproduced from [22]. (b) Target-plasma interface: schematic of velocity distribution of gas particles during evaporation process. Reproduced from [23]. (c) Mechanical coupling: thermal stress: propagation of stress wave in SUS304. Reproduced from [24]. (d) Mechanical coupling: recoil: schematic diagram of melt flow. Reproduced from [25]. (e) Mechanical coupling: shockwave: simulation of Al vapor plume hydrodynamics and thermochemistry using the multi-physics and reactive multi-phase computational fluid dynamics (CFD) code HyBurn. The initial state of the Al vapor plume is a sphere with a temperature of 20000 K and a radius of 0.25 mm. Reproduced from .[26]

### *2.1 Laser ablation*

Laser ablation is an important stage in the formation of laser-induced plasma, determining the initial composition and concentration of the plasma plume. When a pulse laser irradiates a solid substance, the photon energy is absorbed by the electrons in the material and converted into electronic excitation energy, further being transferred to the lattice through electron-phonon interactions [27]. The thermal effects of laser energy deposition will lead to melting, evaporation, and phase explosion on the surface of the target material, along with possible mechanical effects such as thermal expansion, elasticity and plastic stress within the target [28]. All of these processes rely on the interaction between the laser and the target, where factors such as laser wavelength (UV, Vis, IR), irradiance (MW/cm$^2$-GW/cm$^2$), pulse width (fs, ps, ns), target material (metal, organic compound, semiconductor), and environment (vacuum, ambient gas) significantly influence the progress of laser ablation.

### *2.1.1 Normal boiling and vaporization*

During the duration of nanosecond pulse laser, the electrons and lattice inside the material are in equilibrium. The spatio-temporal evolution of the temperature field inside can be described using the Fourier heat conduction equation. If the irradiance of the pulse laser is higher than the material ablation threshold, ablation occurs on the target surface. The mechanism for material removal from the target varies from surface vaporization (surface mass removal) to phase explosion (volume mass removal) in different ranges of laser irradiances. The models based on normal melting and evaporation mechanisms are commonly used to determine the ablation threshold of target materials and to predict crater morphology, which are critical for laser processing, laser cleaning or surface treatment, etc.

The ablation threshold refers to the laser power density required for the onset of mass removal phenomenon under focused laser irradiation. Under different targets and laser parameters, the ablation threshold will undergo significant changes. Li *et al* established a thermal field model for nanosecond laser cleaning of Ti alloys with surface-adhered oxide films to study the ablation behavior at different laser irradiances (laser: 1064 nm, 100 ns, 6.37-50.95 MW/cm$^2$, 0.5 mm, 50 kHz; target: TA15 Ti alloy) [29]. Their results show that with increasing laser irradiance, the temperature of the oxide film will exceed its boiling point, and the ablation threshold of the oxide film is determined to be 19.11 MW/cm$^2$. If the laser irradiance is further increased on the basis of 31.85 MW/cm$^2$, the peak temperature in the target will exceed the ablation threshold of Ti alloy, leading to crack formation on the target surface as also observed in experiments. Precise material properties are required to consider the ablation threshold obtained from the model solution as reliable, however, the temperature dependence of material thermophysical and optical properties is often overlooked. Tao *et al* found in simulating the interaction of multi-pulse nanosecond laser with silicon that the model predictions were lower than the experimental values (laser: 1064 nm, 200 ns, 0.8 mJ/pulse (maximum); target: Si work piece; atmosphere: air) [30]. This discrepancy arises because the optical absorption coefficient of silicon has a strong dependence on temperature and material phase, leading to significant differences in residual heat effects in the model with/without considering its dependence. In addition, Starinskiy *et al* considered environmental factors and studied the differences in ablation threshold of targets in liquid, air and vacuum (laser: 1064 nm, 9 ns, 0.05-6 J/cm$^2$, 0.5-1 mm$^2$; target: Tin; atmosphere: vacuum (2 Pa)/air/deionized water) [31].

Crater morphology is the direct observation object in experiments, such as crater diameter, depth and profile shape. Its simulation plays a positive role in understanding laser ablation and explaining experimental phenomena, but it imposes higher requirements on the accuracy and dimensionality of the simulation. Peng *et a*l established a three-dimensional melting and vaporization heat transfer model to simulate the laser ablation process of A6061 aluminum alloy without considering the fluid flow of the liquid phase (laser: 355 nm, 6 ns, 0.5-2.0 J/cm$^2$, 0.7 mm$^2$; target: AA6061 Al alloy) [32]. Their model shows errors controlled below 5% compared to experiments when predicting the expansion of ablation depth and width under laser energy densities of 0.5-2 J/cm$^2$ and laser shots of 1-100. Wang *et al* used a similar model as shown in figure 2. The model predicted the width and depth of grooves formed by scanning laser processing of aluminum alloy and it also matched well with experimental results (FEM; laser: 40 μm, 7 ns, 600 kHz; target: Al alloy) [33]. Additionally, Acosta *et al* considered the capillary melt flow behavior of the target when ablated to a liquid state in a two-dimensional laser ablation model. This model more accurately simulates the molten pool and teeth regions on the ablated target surface (laser: 532 nm, 1 ns, 0.82-0.85 J/cm$^2$; target: Si) [34].

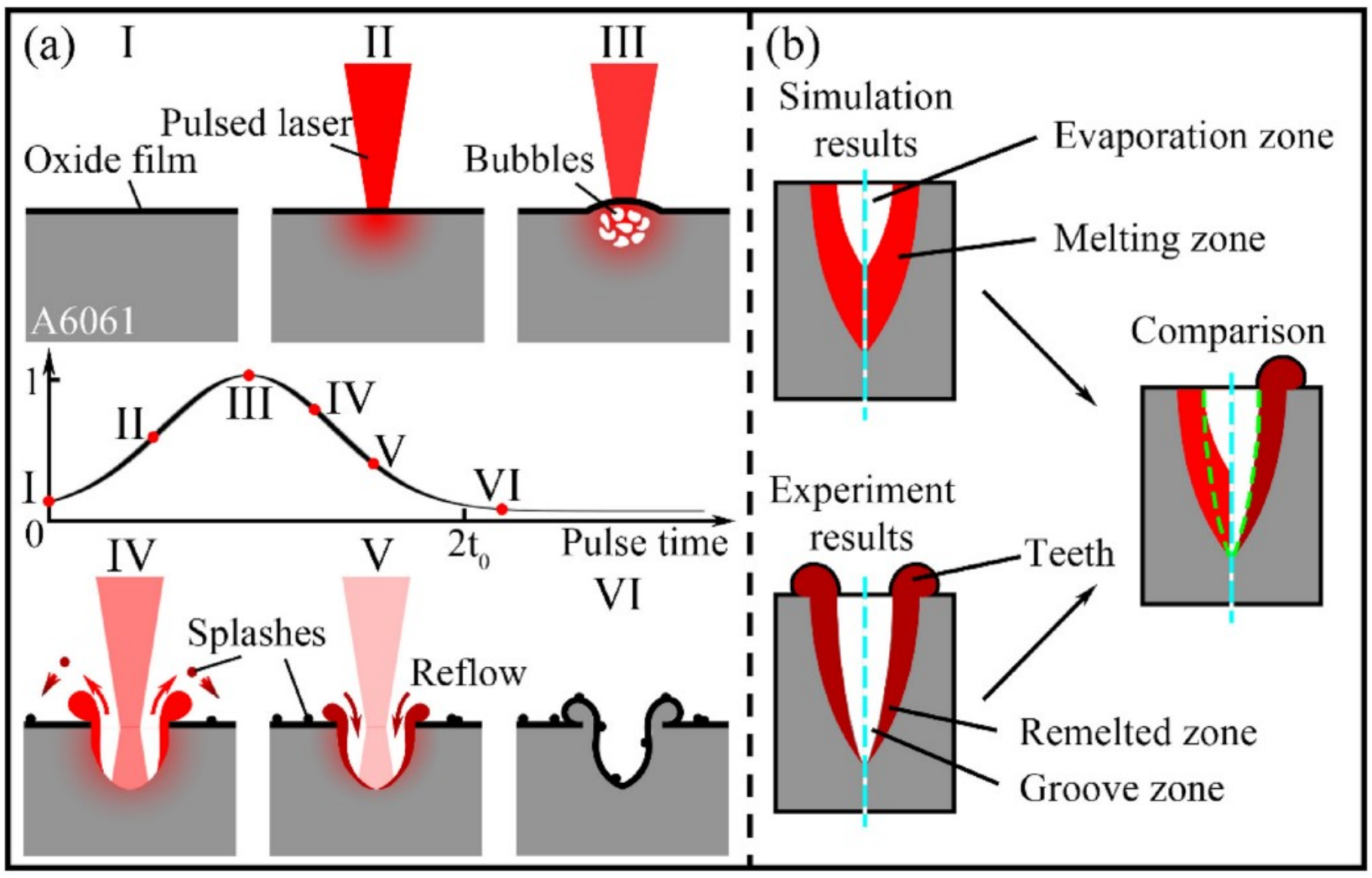

**Figure 2.** (a) Micro-texturing formation mechanism; (b) Schematic diagram of the comparison between the experimental and simulation results. Reproduced with permission from [33].

### *2.1.2 Phase explosion*

Under high-power laser irradiation, the outer layer of the target material can be heated to near the critical temperature ($T_c$), specifically 0.8-0.9 times the $T_c$ [35]. In such conditions, explosive ejection of liquid melt and droplets occurs and phase explosion becomes the predominant mechanism of mass removal. Significant density fluctuations at high temperatures facilitate the formation of vapor bubbles within the liquid phase. The nucleation rate of the liquid phase target will increase sharply when the target surface temperature $T_s$ approaches 0.9 times the $T_c$. This leads to a rapid transition of the target from liquid phase into a mixture of droplets and vapors, with rapid removal of the target surface mass. This ablation mechanism is referred to as phase explosion. The uniform nucleation model has also been employed in simulating laser ablation of nanoparticles [36].

The mechanisms of ablation under high-power laser are contentious, including uniform nucleation phase explosion, non-uniform nucleation, supercritical overheating of liquids and dielectric transition phenomena. Autrique *et al* proposed a mechanism for volumetric mass removal in the supercritical region during high-power copper ablation, distinct from surface vaporization and phase explosion (laser: 532 nm, 6 ns, 4.5-7.5 J/cm$^2$, 1.3 mm; target: Cu; atmosphere: Ar) [37]. This mechanism suggests that the outer layer temperature can reach $T > T_c$, leading to the disappearance of surface tension at the liquid-vapor interface. The volumetric mass removal involves treating regions of the target that exceed the critical temperature $T_c$ as dense weakly ionized plasma, thereby increasing the ablated mass on the target surface. Compared to models that only consider surface mass removal, the volumetric mass removal mechanism reduces the error between the model and experimental observations. Galasso *et al* noted that computed target temperatures exceeding the critical temperature were due to the use of constant thermal properties above the melting point, which deviates from physical scenarios (laser: 532 nm, 10 ns, 1.8 mm; target: Si; atmosphere: air) [38]. Consequently, they embedded a real equation of state (EOS) and a phase transition double-node curve into their model to correct the results. In fact, they also considered the liquid phase region reaching the critical temperature $T_c$ as dense plasma. Zhang *et al* held a different view for simulating and explaining the phase explosion mechanism. They believed there was an intermediate layer between the ablation layer and the liquid layer during the ablation process. This is due to the dielectric transition caused by the dielectric properties of the metal target when the dielectric layer temperature reaches 0.8 times the $T_c$. At this point, the high reflectivity of the metal layer and the high transparency of the dielectric layer cause more laser energy to be deposited on the target surface, further forming phase explosion (laser: 1064 nm, 5 ns, 2-10 J/cm$^2$; target: Al alloy; atmosphere: air (10$^{-3}$-10$^5$ Pa)) [39].

### *2.1.3 Thermal coupling of target-plasma parameters*

Laser ablation models establish links between material physical parameters and experimental characteristics. If differences in experimental phenomena are solely caused by specific physical parameters of the same target material, a calibration relationship between ablation characteristics (crater volume, depth and so on) and that particular physical parameter can be established, thereby achieving inverse measurement of that parameter to some extent. Morozor *et al* computed the ablation processes of various materials based on a thermal model of laser ablation, deriving analytical formulas for effective evaporation time and average surface temperature during evaporation (one-dimensional non-stationary heat conduction equation; laser: 10 ns, 0.1-8 J/cm$^2$; target: metals) [40]. This analytical relationship provides a method to directly obtain peak surface temperature of the ablation target by measuring the evaporation depth. Yang *et al* measured transient surface temperatures during sliding friction of high-speed train materials based on the influence of sample surface temperature on ablation amount and LIBS spectra (laser: 1064.07 nm, 6 ns, 50 mJ; target: brass; atmosphere: 20/100/200/300 ℃ initial temperature) [41]. Their results show an increase in initial surface temperature of the target which leads to faster melting and evaporation. Therefore, a calibration relationship between initial target surface temperature and ablation amount may be established. It still needs to be done with caution although this approach offers a potential method to measure material physical parameters. Since the measured ablation characteristics are the result of the coupling of various influencing factors, focusing solely on the impact of a single target material property on the ablation process may be inaccurate. In fact, most target material property parameters are temperature-dependent. The change in a single parameter will inevitably affect other material property parameters. In such cases, the calibration relationship between material properties and ablation characteristics may be affected by disturbances from other parameter changes.

## *2.2 Mass and energy transfer from target to plasma*

### *2.2.1 Component change*

The essence of evaporation is the transfer of mass and energy from the condensed state of the target to the phase space. Mass transfer can be described by the evaporation flux. For single-component materials, the evaporation flux is obtained from the Hertz-Knudsen equation. However, for multi-component targets, it is necessary to consider whether the ablation of the target satisfies stoichiometric balance. If the differences in melting and boiling points of alloy components are small, the evaporation flux ratio of each component can be considered the same as the stoichiometry of the sample elements. Ait Oumeziane *et al* suggested that the compound TiC forms equal molar fractions of Ti and C atoms after ablation, but this has not been experimentally verified (the LCPFCT (Laboratory of Computational Physics Flux Corrected Transport) algorithm; laser: 248 nm, 20 ns, 0.93 J/cm$^2$ for Ti & 17.5 J/cm$^2$ for TiC; atmosphere: He (10Pa)) [42]. Morozov *et al* suggested that the ablation of gold and silver alloys satisfies stoichiometry, but they also consider the earlier evaporation time of silver due to its low boiling point. The difference of unit evaporation fluxes between different components of gold and silver is simulated by introducing delayed evaporation time (laser: 532 nm, 7 ns, 6-20 J/cm$^2$; target: Au/Ag/Au-Ag alloy; atmosphere: vacuum) [43].

It is commonly believed that the ablation of most alloys does not satisfy stoichiometric balance when the effects of fractional distillation caused by differences in melting and boiling points cannot be ignored. For example, in Cu-Zn alloys, the Zn component has a higher concentration in the LIP due to its lower boiling point [44]. Klassen *et al* introduced the activity coefficient $\gamma$ into the calculation of the evaporation flux [45]. Their corrected multi-component evaporation model can correct the composition of the ablation plume and thus improve the quantitative analysis ability of LIBS for alloys. If the activity coefficient $\gamma$ for each component is equal to 1, it degenerates into stoichiometric balance. In general, $\gamma$>1 corresponds to the lower boiling point element components in the alloy, while $\gamma$<1 corresponds to the higher boiling point element components, and the activity coefficient is also temperature-dependent [46]. The lack of activity coefficients is the main difficulty in simulating multi-component systems. Activity coefficients can be fitted from experimental data [47] or calculated using Raoult's law. The work in [45] provided activity coefficients for the binary mixture system Al-Ti. Wang *et al* used this binary activity coefficient model to determine the Al-Ti evaporation flux of the alloy Ti-6Al-4V alloy during ablation, which matched well with the experiment (CW laser: 6-18 kW, 50 μm; target: Ti-6Al-4V; atmosphere: vacuum) [48]. Tao predicted the activity coefficients of ternary mixtures using the molecular interaction volume model (MIVM) method [49].

The stoichiometry of deposited films is the most important parameter determining the coating performance in pulsed laser deposition (PLD). However, it is not easy to maintain the stoichiometric ratio of the target under real experimental conditions. Acquaviva *et al* found that the stoichiometry of deposited films is influenced by the atomic mass. This is because species with different atomic weights have different angular distributions during evaporation (laser: 308 nm, 30ns, 7/3.5/2.5 J/cm$^2$, 1.3 mm$^2$, 10 Hz; target: $Co_{67}Cr_7Fe_4Si_8B_{14}$; atmosphere: air ($10^{-5}$ Pa)) [50]. Itina *et al* suggested that the stoichiometry of deposited films is also affected by the substrate adsorption probability and the energy of deposited particles [51]. It should be noted that the stoichiometry of deposited films is not entirely consistent with the stoichiometric evaporation flux discussed in this section. PLD is obviously influenced by two processes instead, one is the multi-component evaporation process, and the other is the deposition process of evaporated particles on the substrate. Therefore, the combined effects of these two processes should be considered together for multi-component simulations of PLD.

*2.2.2 Escape velocity distribution*

The velocity distribution of escaping particles affects the energy transfer to the phase space during evaporation. From a numerical simulation perspective, the boundary conditions at the target surface (i.e., the capture and continuity assumptions of the gas-liquid phase change interface) are crucial for simulating the energy transfer during laser ablation. Zhao [52] and Peng [53] used an interface grid treatment method based on the free surface flow assumption to simulate laser cutting of metal sheets and laser-induced bubble formation. This method assumes that the particle velocity at the interface is equal, and it models the surface motion and pressure boundary conditions by solving mass and momentum fluxes across the interface [54]. It is similar to the volume of fluid method. In addition, Knight described the transition of evaporated particles from the thermodynamic and kinetic non-equilibrium on the liquid side of the target to the Maxwell-Boltzmann equilibrium distribution on the plume side using jump conditions [55]. These jump conditions describe the temperature, pressure and density of the vapor based on the Knudsen layer hypothesis. It ensures a reasonable connection between the target and the plume domain. Gusarov *et al* focused on the Mach number during laser ablation and the variation of condensation flux, and rigorously analyzed the gas dynamic condensation conditions during laser ablation (laser & target: 193 nm, 12ns, 5.3 J/cm$^2$ for Al; 266 nm, 6ns, 3.5 J/cm$^2$ for Au) [56]. They suggested that subsonic or sonic evaporation occurs during the laser pulse duration, while subsonic reverse condensation and sonic condensation occur after the pulse. It is also worth noting that when considering the laser absorption of ionized plasmas, the decrease in saturation pressure occurs more rapidly and the condensation occurs before the laser pulse stops.

*2.3 Mechanical coupling of target-plasma-atmosphere parameters*

When the laser irradiance exceeds $10^8$ W/cm$^2$, the mechanical transfer of energy in laser-target interactions needs to be considered [57]. There are three sources of mechanical energy transfer here:

(i) in-target: Laser energy absorbed in solid samples leads to rapid heating and volume expansion, generating significant stress, fractures, and plastic deformation internally. Khosroshahi *et al* associated the initial pressure generated with the thermal expansion coefficient, the speed of sound, and the specific heat capacity for quantitatively characterizing the thermoelastic stresses generated by laser-induced ultrasound (laser: 1.06 μm, 7 ns, 0.6-200 J/cm$^2$, 1 mm; target: teeth) [58].

(ii) target-plasma: Rapid detachment of target material ablation products from the target surface generates recoil and stress pressure on the surface. Shannon *et al* estimated that a medium-focusing beam with a diameter of 100 μm can produce stresses on the target surface ranging from $10^9$~$10^{10}$ Pa/$10^4$~$10^5$ bars [57].

(iii) plasma-atmosphere: When a gas medium exists above the target, high-speed evaporating particles leave the target surface and collide with the background gas to accelerate it. When their velocity exceeds the local speed of sound, shockwaves are generated, which, in turn, confine the expansion of the plasma plume [59, 60].

Many phenomena demonstrate that these mechanical energy transfers are strongly coupled with various aspects of nanosecond pulse-target interactions, profoundly affecting target ablation, expansion, ionization, and even radiation processes. Non-exclusive examples include LIBS, which has been shown to have the potential to characterize the mechanical properties of material surfaces, such as hardness [61], surface stiffness [62], specific characterization indicators including ratios of ion to atomic spectral line, plasma temperature, electron number density, ablation mass, and etc. [63-65]. In Laser-Induced Plasma-Assisted Ablation (LIPAA) for micro-structuring transparent material surfaces, the morphology of crater is strongly influenced by shockwaves and recoil pressure [66]. High-frequency and broadband ultrasonic waves generated by laser-target interactions are advantageous for non-contact, non-destructive, remote, and online detection of mechanical structures such as unknown material absorption coefficients, residual stresses, internal defects, and etc [58, 67]. The basic characterization steps of ultrasonic waves are shown in figure 3. Hardening induced by nanosecond laser pulses was found to be essentially the same as that induced by femtosecond laser hardening. [68]. Laser Surface Melting (LSM) can significantly modify metal hardness, roughness, surface/subsurface morphology, and etc [69, 70].

However, these effects remain unresolved to date, not fully explained, nor completely accepted at the fundamental level. Therefore, we re-examined and discussed the modelling of mechanical energy transport in nanosecond pulse-material interactions in this section. Additionally, the mechanical explanations for ultrashort pulse interactions (sub-picosecond and shorter) are also compared concisely in this section.

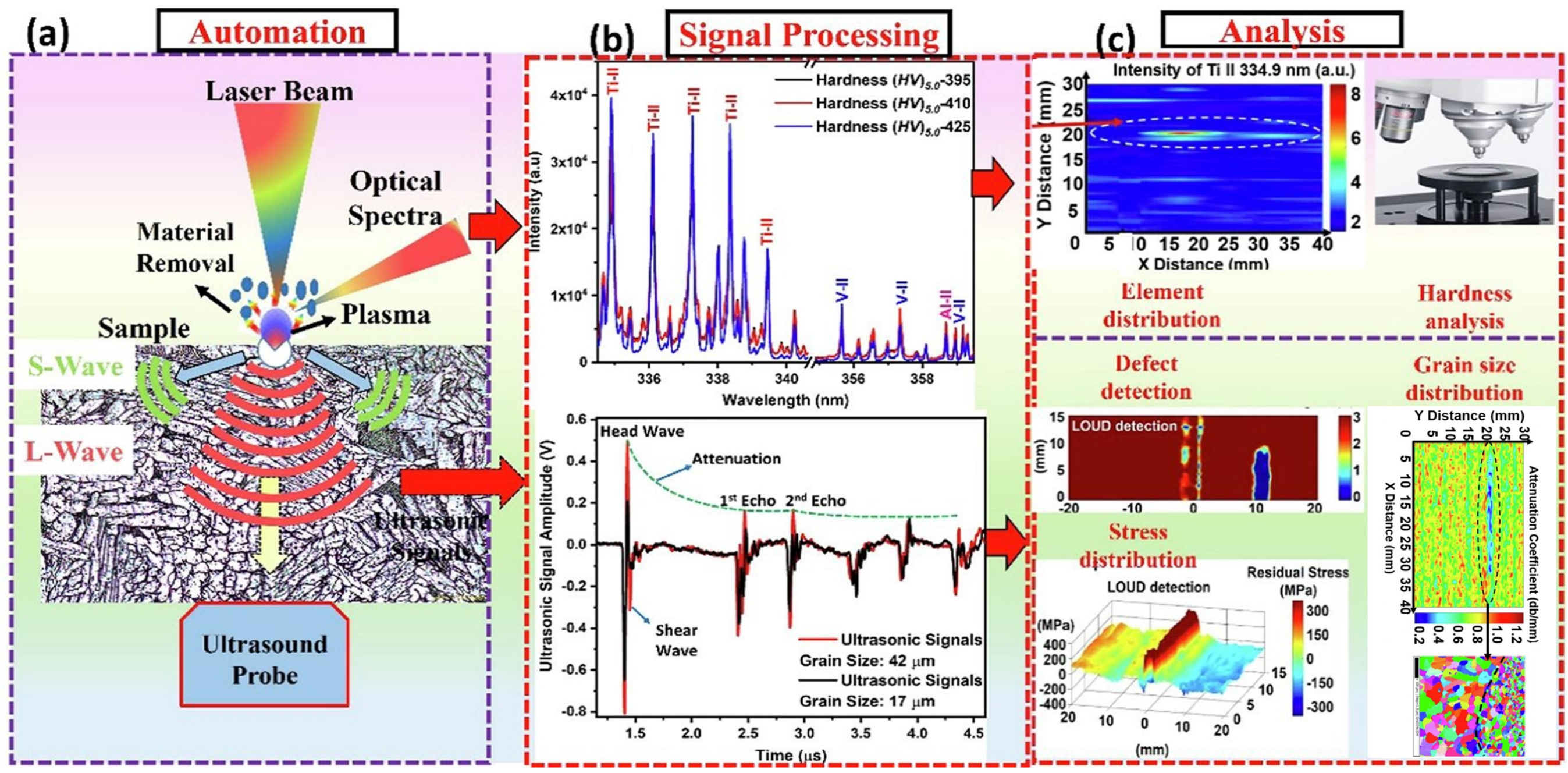

**Figure 3.** (a) Schematic diagram of laser-target thermomechanical effect. (b) Laser ultrasonic signal processing results. The image above shows the optical emission peaks of Ti of different hardness values. Below is the amplitude of ultrasonic signals decreased from the head wave to the respective echoes. (c) The analysis and application of laser ultrasonic signals include several critical aspects: detecting the distribution of elements (Ti element in wire and arc additive manufacturing samples), evaluating the hardness of materials, defecting the detection (detect mapping in welded specimens obtained by laser ultrasound), obtaining three-dimensional residual stress distributions and grain size distribution (spatial distribution of acoustic attenuation coefficient). Reproduced with permission from [67].

### *2.3.1 In-target: thermo-mechanical stress and ultrasonic wave*

Compared to the well-defined, regular boundaries of ablation craters produced by femtosecond lasers [71], Shugaev *et al* observed typically thick and raised rims for ablation by pulsed nanosecond lasers, where the primary mechanism was thermal ablation (laser: 532 nm, 1 ns, 0.6 J/cm$^2$; target: single crystal Si) [72]. These thick rims are formed by the process of the melting and subsequent re-solidification of the material. Laser irradiation results in the generation of sufficiently high tensile residual stress on the target, which subsequently induces cracking and the initiation of cracks in the target material. These phenomena were further investigated in the measurements of Carr *et al* (laser: 355/532/1064 nm, 2.7/4.2/5.5 ns; target: CsI/sapphire/ CaF2/DKDP/fused silica/LiF) [73]. They found that the residual stress is produced by the temperature gradients generated in the space and time domains during laser ablation.

Laser paint cleaning, as one typical techniques for thermal stress delamination involving the thermo-mechanical coupling process between the laser and the paint film [74], mainly includes melting and ablation, thermal stress delamination, and plasma impact [75]. Han *et al* modeled the above three effects in detail based on heat transfer equations and linear thermo-elastic constitutive equations. Taking thermal stress delamination as an example, a high stress difference $\Delta\sigma$ at the interface can be caused by the difference in thermal expansion properties and temperature gradients between the paint film and the substrate. In order to separate the paint film and substrate, the high stress difference $\Delta\sigma$ between the two needs to exceed the adhesion force of the paint film and substrate (laser: 1064 nm, 13.6 ns, 45 μm; target: paint) [76]. The simulation results indicate that three thermomechanical effects correspond to different ablation outcomes. Mechanical action or thermal stress can completely remove the paint film without causing damage to the substrate surface. Melting ablation results in residues on the substrate surface. Plasma shock causes damage to the substrate. Therefore, they suggested that the laser intensity should be kept higher than the thermal stress threshold while lower than the plasma ionization threshold. Lu *et al* reached similar conclusions and specifically analyzed the cleaning threshold of iron-based paint layers (laser: 1064 nm, 10 ns, 0.63-3.5 J/cm$^2$; target: the paint layer and Fe substrate) [77]. In addition, Zhang *et al* analyzed the laser scanning speed using a similar thermodynamic model, suggesting that reducing the laser scanning speed enhances the ablation effect without affecting the effectiveness of thermal stress paint removal (laser: 30 W, 200 ns, 30 μm; target: topcoat, primer and $Al_2O_3$ oxide film on the surface of a 2A12 aluminum alloy substrate) [78].

Laser shot peening (LSP) induces deeper and finer microstructures within the material by inducing more severe plastic deformation, thereby modifying the mechanical properties such as corrosion resistance and fatigue performance [64]. Unlike laser directly focused on the target surface, LSP typically involves covering the target with an absorptive/inert layer to enhance laser absorption while protecting the target from plasma heating, maintaining a longer duration of high plasma pressure on the surface, thus achieving more sustained plastic deformation [79]. Moćko *et al* calculated the spatio-temporal distribution of elastic and plastic waves induced during the plastic deformation process of 304 austenitic steel by laser pulses using the strain rate-sensitive Johnson-Cook equation, where strain rate sensitivity was calibrated through quasi-static and dynamic loading tests (FEM; laser: 1064 nm, 10 ns, 2.5 GW/cm$^2$; target: AISI 304 steel) [79]. Furthermore, compared to an air atmosphere, it is suggested that in a water atmosphere, due to the water layer confinement to the plasma generated, higher magnitude (approximately GPa) recoil pressure will exceed the dynamic yield strength of the workpiece easily, making plastic deformation and residual stress more effectively generated. Based on models of plasma constrained in water and 3D finite element thermal stress, Cao *et al* systematically studied variations in target surface integrity and induced residual stress under single and overlapping water-based LSP occurrences, with the peak residual stress increasing with laser power density, while overlap ratio contributes to peak residual stress without affecting indentation depth [80].

Crack initiation and melting thresholds of materials under extreme heat loads from fusion-related nanosecond pulses have also been studied by coupling thermal and dynamic displacement equations. Besozzi *et al* determined crack initiation thresholds from three aspects, namely (i) when warming up, temperature $T$ increases and reaches the ductile-to-brittle transition temperature (DBTT), (ii) when temperature $T$ continues to exceed the DBTT i.e. $T$>DBTT, and maximum thermal stress exceeds the ultimate stress of the material, (iii) when cooling down at $T$<DBTT, and the maximum thermal stress is higher than the yield strength (laser & target: 1064 nm for bulk W/metallic W coatings, 532 nm for a-$WO_3$ coatings, 7 ns, 3.5-50 mJ/cm$^2$, 0.81 cm$^2$; atmosphere: vacuum) [81].

For certain semiconductors and insulating materials that exhibit strong temperature-activated optical absorption, the laser-supported solid-state absorption frontier model also provides a plausible mechanism for understanding the mechanical damage and energy deposition caused by laser interaction with solid materials (laser: 5/20 ns, 0.25-2 GW/cm$^2$,1 cm; target: polished fused Silica windows) & (0.35 μm, 2.5/8.5/15 ns, 0-25 J/cm$^2$; target: fused Silica) [82, 83]. More specifically, when the laser pulse is focused on the back surface of the sample, the lattice temperature continues to rise due to thermal diffusion. It also forms sub-bandgap defect states capable of absorbing photons and transferring free electrons, which can drive energy deposition. Shen *et al* observed and numerically analyzed the generated sustained absorption waves generated within the solid material, which behave similarly to laser-supported combustion waves in the atmosphere (compressible fluid model; laser: 1064 nm, 7.6 ns, 150 J/cm$^2$; target: fused Silica) [84].

However, in some cases of laser processing, the accumulation of heat and thermal stress must be controlled to prevent the formation of cracks and defects. For such cases, Wang *et al* developed waterjet-guided laser micromachining, in which a high-speed water jet acts as a fiber to transmit the laser beam while removing molten material and accumulated heat, too.

This reduces the thermally affected zone and minimizes thermal deformation and damage to the material [85]. Zhang *et al* simulated transient thermal effects of water-jet guided laser machining on carbon fiber-reinforced plastic (CFRP) materials combining convective cooling boundary conditions of water jets, and further compared the effects of laser duty cycle on heat accumulation and water cooling efficiency (laser: 532 nm, 0.3 μs, 30 W, 30 kHz, 70 μm; target: CFRP composite laminates; atmosphere: water jet-guided laser processing) [86].

### *2.3.2 Target-plasma: recoil pressure*

In addition to thermal elastic relaxation within irradiated samples and mechanical damage induced by high-temperature plasma expansion, the formation of plasma and the rapid expansion-induced shockwave are likely to generate significant recoil pressure at the target-plasma interface, which is also one of the possible reasons for the generation of surface defects and cracks [87]. Depending on the mechanism of nanosecond laser ablation or material removal, the pressure in as well as above the target also changes.

For surface ablation considered as the primary mechanism of material removal, where optical penetration depth is much smaller than the thermal diffusion length, the mechanical coupling between target and plasma is achieved through the Knudsen layer [88, 89]. The plasma plume evaporates forward when the environmental pressure is below the saturation vapor pressure. Otherwise, the plasma plume condenses backward on the target surface. This backward condensation has the potential for applications such as processing porous coatings on the target surface and forming nano/microscale reliefs, and it is also one of the main mechanisms for forming oxide layers on metal surfaces through laser ablation [90]. Gornushkin *et al* gave an approximation for estimating the pressure and deposition ratio of this backward shock, assuming that vapor outside the Knudsen layer escapes at a local speed of sound, with the backward deposition ratio being approximately 0.2 times the total ablated mass, and the recoil pressure being approximately 0.2 times the saturation vapor pressure (laser: 1.7 μm, 100 ns, 1 mJ, 5 J/cm$^2$, 25 μm; target: Ti; atmosphere: $N_2$:$O_2$:Ar = 78:21:1) [91].

When the target is heated to a higher temperature state, material removal tends to push the target into a metastable region of liquid-vapor mixture. At this point, the process involves nucleation and growth of bubbles in the surface liquid layer, and the target-plasma interface simultaneously experiences liquid pressure $P_l$, bubble pressure $P_{\mathrm{bub}}$, and ambient pressure $P_{\mathrm{amb}}$. Autrique *et al* proposed a feasible method for solving this condition through controlling $P_l$ and $P_{\mathrm{amb}}$ with bidirectional lines on the phase diagram, combining it with bubble nucleation theories (Volmer-Doring theory and Eotvos rule) to solve for bubble growth and collapse (laser: 532 nm, 6 ns, 10 J/cm$^2$, 1.3 mm; target: Cu; atmosphere: Ar) [37].

As the surface temperature continues to increase until reaching the thermodynamic critical temperature $T_c$, the surface tension vanishes as the liquid-vapor interface disappears. Once the target cells arrive in their supercritical state, the target surface should be repositioned. At this point, supercritical cells are introduced into the vapor domain and treated as the weakly ionized plasma.

In addition to the recoil pressure on the target surface, for objects with higher requirements on processing precision and texture fineness, such as artificial bone implant materials [92], further consideration should be given to the effects of interfacial fluid mechanics. Zhao *et al* discussed these effects in detail, including the Marangoni effect caused by surface tension gradients at the gas-liquid interface. and momentum loss within the slurry region where gas and liquid coexist (also known as Darcy resistance) (laser: 355 nm, 50 ns, 22.9 J/cm$^2$, 35 μm; target: Ti alloy) [93].

Singh *et al* investigated an interesting phenomenon. Their experimental studies show the laser ablation rate and crater depth are increased when an external magnetic field is present. They tried to explain this by considering the magnetic pressure in addition to the recoil pressure and the saturation vapor pressure (laser: 532 nm, 10 ns, 1.4 GW/cm$^2$, 300 μm; target: Cu; atmosphere: magnetic field (0.5 T)) [94]. However, their preliminary calculation indicates that under 0.5T transverse magnetic field, the recoil pressure borne by the target shows almost equal to that without a magnetic field. At this point, the experimental increase in ablation is attributed to the thermal conductivity changes affected by the magnetic field.

Liquid-immersion nanosecond pulsed laser micromachining is proposed to address thermal losses, redeposition, and condensation induced by long pulse durations. Mak *et al* utilized a one-dimensional expansion recoil approximation and estimated that the recoil pressure in water medium $P_{H2O}$ (9.35 kbar) is much greater than that in an air atmosphere $P_{air}$ (1.11 kbar) (laser: 349 nm, 4 ns, 1 kHz; target: GaN; atmosphere: deionized water) & (laser: 1.06 μm; target: Cu; atmosphere: vacuum) [95, 96]. They concluded that for the condition of the laser energy maintained constant during one pulse duration $\tau$. The water confinement helps in rapid detachment of plasma from the surface and reducing redeposition.

Furthermore, the mechanical coupling of laser-plasma-target interaction exhibits a strong wavelength dependence. Taking laser propulsion as an example, Phipps *et al* defined a mechanical coupling coefficient $C_m$ as the ratio of the total momentum imparted to the target by laser ablation ($I_m=mv$) to the incident laser energy ($e=I_{L\tau}$). They proposed a special type of empirical formula to elucidate the correlation between the maximum $C_m$ and laser wavelength $\lambda$. This empirical formula is applicable in vacuum, single pulse, incident laser energy greater than the threshold of plasma formation. It has been experimentally verified on five laser systems with aluminum alloy and C-H type materials [97]. After that, Yuan *et al* discussed the dependence of mechanical coupling coefficient $C_m$(Al) on $\lambda$ in vacuum through a one-dimensional (1D) bulk absorption model simulating the solid target ablated by the laser [98]. The numerical results were compared with Phipps' empirical curve to verify the dependence of $C_m$ on $\lambda$.

### *2.3.3 Plasma-atmosphere: shockwave expansion*

As one of the sources of momentum at the target-plasma interface, high-temperature and high-density plasma, once formed, simultaneously impacts the target surface from above and expands upwards to compress the surrounding cool gas atmosphere. When the plasma's expansion velocity exceeds the local speed of sound, shockwaves are generated and driven in the atmosphere. Compared to cylindrical shockwaves generated by femtosecond/picosecond pulses [99], Zeng *et al* observed a more regular semi-spherical shape for nanosecond pulse-induced shockwaves (laser: 266 nm, 3 nm, 220 μJ, 11 J/cm$^2$; target: Si; atmosphere: air) [100]. Pan *et al* and Zhao *et al* gave one explanation that shorter pulses tend to pre-ionize the air through multiphoton, tunnel, and avalanche ionization when propagating in air, forming an initial air plasma channel that affects the expansion of the target plasma shockwave (laser: 800 nm, 50 fs; target: Si; atmosphere: air) & (laser: 800 nm, 100 fs, 1 kHz; target: Si) [101, 102]. In contrast, the trailing part of the nanosecond pulse is absorbed via electron-neutral, electron-ion inverse bremsstrahlung, and photoionization, which typically results in an initial ns-plasma that can maintain a high radial expansion velocity, higher electron temperature, and more electron number density compared to ps/fs-plasma. However, it should be noted that the high-temperature state of the plasma does not necessarily have to be reached by absorbing longer timescales of laser radiation. This is due to the fact that the optical and thermal properties of the target, the continuous radiation and the thermal convection losses into the cooling atmosphere also strongly influence the plasma state [103].

The Taylor-Sedov explosion theory provides a "middle-field" self-similar solution for shockwave propagation [104, 105]. The explosion mass and energy are defined as $M_0$ and $E$ respectively, while the undisturbed gas density and pressure are designated as $\rho_0$ and $P_0$ respectively. The characteristic size of the explosion source can be described as $(3M_0/2\pi\rho_0)^{1/3}$, and the premise of Taylor-Sedov explosion theory is that $(3M_0/2\pi\rho_0)^{1/3} \ll R \ll (E/P_0)^{1/3}$ [106]. It can be demonstrated that the propagation distance $R$ of the explosion wave needs to be significantly greater than the characteristic size of the explosion source. At the same time, the undisturbed pressure of the wave front $p_0$ must be sufficiently low that $(E/P_0)^{1/3}$ is considerably greater than the propagation distance $R$. Therefore, the Taylor-Sedov model can be applied only when the characteristic size of the explosion source and the undisturbed pressure $P_0$ can be ignored. The model has been shown to have a remarkable predictive effect in simulating the evolution of laser-induced shock waves at low pressure [107-109].

However, under atmospheric pressure, the particles at the plasma front undergo frequent collisions with air molecules, continuously losing speed until the front pressure equals the ambient pressure, causing the plasma to stop [110]. Geohegan *et al* and Min *et al* utilized another drag model assuming that the plasma experiences viscous forces proportional to its expansion velocity (laser: 248 nm, 3.8 ns; target: YBCO films; atmosphere: vacuum/oxygen (100 mTorr)) & (laser: 1064 nm, 10 ns, 300 mJ; target: Si; atmosphere: vacuum) [111, 112]. Additionally, Chen *et al* provided a set of equations through continuity requirements to completely describing the shockwave propagation process, including the initial formation, growth, and free decay stages of the shockwave (laser: 1.06 μm, 15 ns, 40.5/105 mJ, 50 μm; target: Al) [106].

Furthermore, apart from the shockwave existing between the plasma and the atmosphere, shockwaves are also found within the rapidly expanding plasma itself, termed as the internal shockwave for the purpose of differentiation [113, 114]. These internal shockwaves reflect back and forth within the plume, considered as an inevitable consequence of the balance between pressure and velocity continuity conditions. Wen *et al* found that the front of these internal shockwaves becomes distorted from a semi-spherical shape during impact with the target surface due to single-shot impacts and reflections (iterative comparison of the simulated (solving the integrated conservation equations of mass, momentum, and energy) and experimental trajectories of the external shock wave and contact surface; laser: 1064 nm, 4 ns, 300 μJ; target: transition metals) [115]. As the plume approaches the surface at a higher velocity gradient, shear stresses due to surface viscosity lead to the formation of a vortex ring near the surface, providing additional outward velocity to the internal shockwave, causing it to become concave [116].

Apart from near-surface viscosity effects leading to observable vortex rings, high-Z elements with their relatively heavy atomic masses and complex atomic structures, often exhibit more complex chemical reactions and evolution behaviors during plasma formation [117]. Harilal *et al* related the complex morphology of shockwaves induced in high-Z plasma under different oxygen partial pressures to turbulent transport in low-Z environments. Specifically, the presence of Rayleigh-Taylor and Richtmyer-Meshkov instabilities at the plume edge complicates the modelling of shockwaves in high-Z plasma (laser: 1064 nm, 6 ns, 12 J/cm$^2$; target: U; atmosphere: vacuum) [118].

The free expansion of plasma in vacuum differs significantly from its behavior in the atmospheric medium. A notable feature is that shockwaves are not allowed to approach the vacuum region; instead, a smooth rarefaction wave takes place, having a fan-shaped profile bounded by two boundaries corresponding to the head of the rarefaction wave and the plasma contact, and the tail of the rarefaction wave merging with the vacuum front. Oderji *et al* reconstructed the rarefaction wave formed by tungsten in a vacuum environment using variable-sized grids (laser: 1064 nm, 7 ns, 131 J/cm$^2$, 300 μm, 20 Hz; target: W; atmosphere: vacuum) [119]. The detailed differences between vacuum and air laser-induced breakdown are also explained in the review of Giacomo *et al* [120].

Besides gas medium, Wu *et al* also studied the propagation characteristics of laser-induced shock waves in water. The water pressure has a greater impact on shock waves compared to gas atmosphere. The results indicate that the residual stress and mechanical damage generated by laser ablation on the target surface cannot be ignored, as this mechanical action significantly affects the propagation characteristics of the shock wave (laser: 1064 nm, 8 ns; target: Al; atmosphere: pure deionized water) [121]. Furthermore, the specificity of the liquid environment lies in the need to address the dynamics of cavitation bubbles and their interaction with rigid wall surfaces. Zhang *et al* provided a detailed review of the application of laser-induced cavitation in liquid media for the material surface processing such as metal strengthening, cell perforation, and micro-forming techniques replacing micromachining [122].

Vogel *et al* observed the shockwave front detaching from the plasma and the expansion velocity of the shockwave to be almost identical to the bubble wall velocity during its initial formation. However, the expansion speed of the shock wave and the bubble wall are not comparable (the Gilmore model of cavitation bubble Dynamics; laser: 1064 nm, 6 ns/30 ps; target: an ophthalmic contact lens; atmosphere: distilled water) [123]. Other early studies presumed a spherical symmetry of shock waves and bubbles, with the Taylor-Sedov point explosion and Gilmore model being used to describe the motion of shock waves and bubbles, respectively [124]. However, non-spherical evolution has been discovered, attributed to the Bjerknes effect of bubbles near rigid wall surfaces. Lu *et al* proposed an extension of the Taylor-Sedov model with the analytical form of the propagation path of a rotating ellipsoidal shockwave, and demonstrated the consistency with experimental recorded shockwave profiles (laser: 1.064 μm, 24 ns, 4 J, 0.1 mm; target: deionized water; atmosphere: air) [125]. A more comprehensive model developed by Lechner *et al* supplements the importance of liquid viscosity for the dynamics of bubble shape, although the viscosity of water previously being considered negligible (bubble model, finite volume method (FVM)) [126]. More detailed development of the non-spherical model of cavitation bubbles can be found in the review [122].

In addition, the interaction between nanoscale particles and nanosecond pulses inducing shock waves has also been investigated. Soumyashree *et al* found that with increasing background pressure (from $<5\times10^{-2}$ mbar to 1 mbar to 1000 mbar), the *r-t* diagrams of shockwave expansion induced by LIP with nano-particles follow different descriptions changing from adiabatic free expansion, Taylor-Sedov strong explosion, to the drag model, respectively (drag model; laser: 1064 nm, 7 ns, 60/90/120 mJ; target: Cu and Al-6061 alloy; atmosphere: 0.05 mbar air, Ar gas at 1 mbar and air at 1000 mbar) [127]. Compared to cases without nano-particles present, the initial expansion velocity of shockwaves in LIP with nano-particles is higher. Additionally, Yu *et al* provided the theoretical form of the compressive force of the shock wave. It can test the feasibility of directing microspheres propelled by shock waves induced by laser-induced breakdown of air (laser: 532 nm, 10 ns; target: glass microspheres) [128].

It should be noted that shock waves can also be modelled indirectly by coupling the flow field and the equation of state [129]. The states of plasma play an important role in coupling the flow field. Casavola *et al* [130] studied the influence of chemical reactions on plasma expansion under considerations of local thermodynamic equilibrium (LTE), chemical non-equilibrium (CNE), and free flow approximation (FFA) without chemical reactions. Their results showed that shockwave expansion velocity is slowest under LTE; free flow approximation results in the fastest shockwave velocity, while the presence of chemical reactions slows down the shockwave expansion and reduces the plasma temperature (laser: 248 nm, 30 ns, 5 J/cm$^2$; target: Ti; atmosphere: $N_2$ ($10^{-3}$-$10^5$ Pa)).

The last simple yet potentially significant regularity discovered in the interaction between picosecond/femtosecond pulses and materials is that the volume of ablation craters produced shows inversely proportional to the material's bulk modulus/Young's modulus. Bigoni *et al* and Gamaly *et al* interpreted it as a work done by the shockwave gradually attenuating into acoustic waves as it propagates within the target. They both related the effective stopping distance of the shockwave to the pulse energy $E_{abs}$ absorbed by the material internally through the bulk modulus/Young's modulus (laser: 532 nm, 50 ps, 600 MW; target: Al, Si, $SiO_2$; atmosphere: air/vacuum) [131, 132]. These findings hold true for a large number of materials (Al, Si, $SiO_2$, sapphire, fused silica, polystyrene), and link the laser ablation characteristics to the thermal and mechanical properties of materials, once again demonstrating the inevitability of mechanical coupling between target-plasma-atmosphere, albeit being discovered in ultra-short pulses. Therefore, further attention to the mechanical coupling under the action of nanosecond pulses on materials needs to be emphasized.

## 3. Nanosecond laser-plasma interactions and plasma radiation

When the irradiance of the incident laser exceeds the breakdown threshold of the plasma, the plasma rapidly forms and undergoes complex physical and chemical processes, including expansion, ionization, recombination, and radiation in the vacuum or gaseous medium in figure 4. In Section 3.1, advances in models describing the fluid expansion behavior of the nanosecond LIP are presented, focusing on transport phenomena, radiation losses, magnetic field effects (which are often neglected in nanosecond laser-plasma interactions, and here their potential importance being reiterated), and particle simulations combined with the advantages in modeling dilute gas flows. In Section 3.2 the rapid ionization-recombination processes in the nanosecond LIP are discussed, which determine the relative abundance of various charge state ions, atoms, molecules and etc., including an exploration of population distributions, several physical explanations for the ion acceleration phenomenon, and a preliminary discussion of the plasma chemistry. In Section 3.3 a review of recent advances in atomic and molecular emission in the nanosecond LIP is provided since plasma radiation of great importance as a direct experimental observable for spectral analysis and spectral-based elemental analysis techniques.

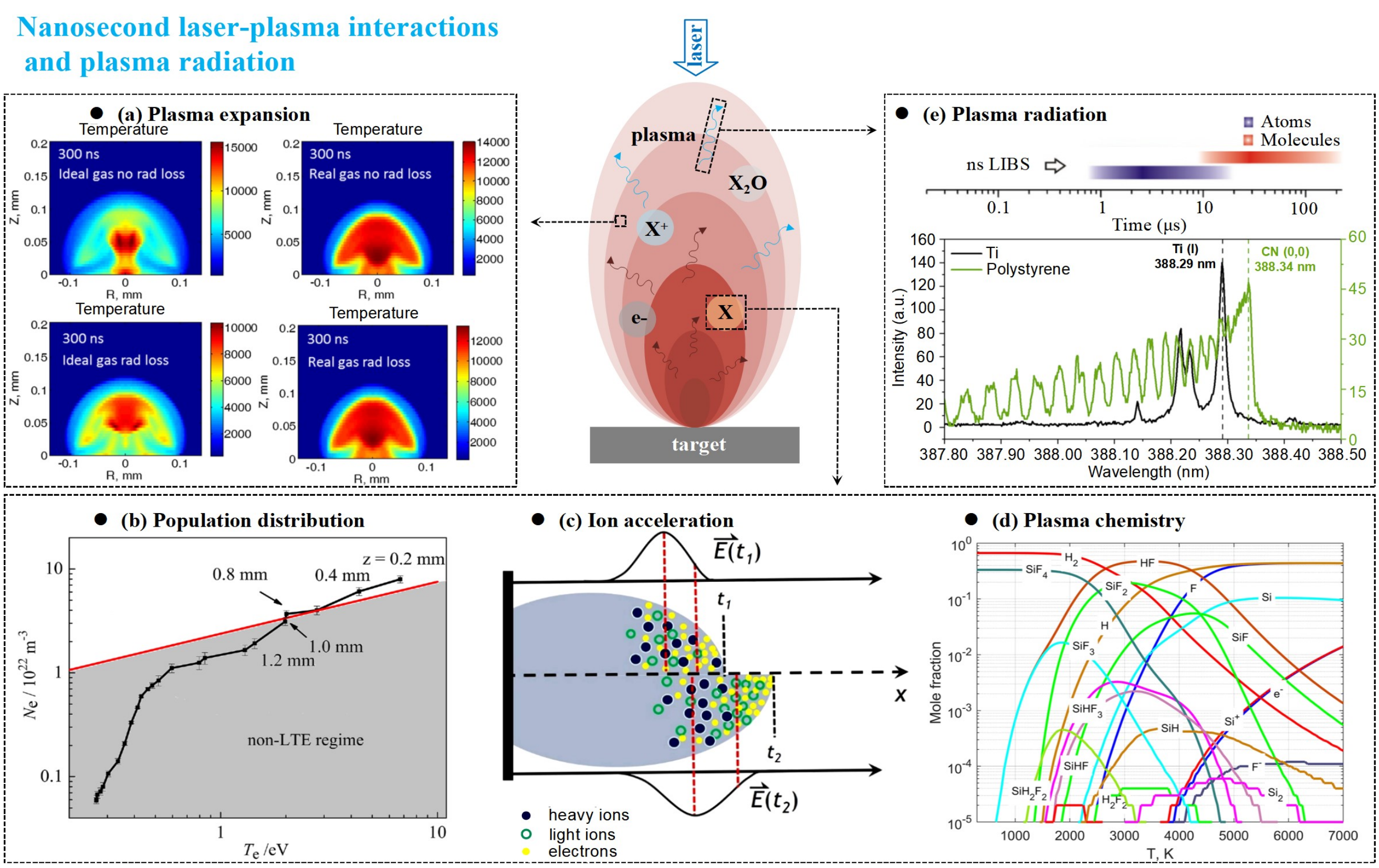

**Figure 4** Schematic diagram illustrating the progress of nanosecond laser-plasma interactions and plasma radiation. Reproduced from [133]. (a) Plasma expansion: the temperature fields at 300 ns (left column: the ideal gas plasma with (bottom) and without (top) radiative losses; right column: the same for the real gas plasma). The initial plasma parameters are $r_0/z_0 = 2/3$, volume $V_0 = 3.8\times10^{-5}$ cm$^3$, mass $M_0$ = 100 ng, and energy $E_0$ = 5 mJ so that the ellipsoid semi-axes are $r_0$ = 230 μm and $z_0$ = 340 μm. The plasma is composed of 95% Al, 4% Si, and 1% Mg and expands into an ambient gas (argon) at the atmospheric pressure and temperature 300 K. Reproduced from [134]. (b) Population distribution: log–log plot of $N_e(z)$ versus $T_e(z)$ derived assuming a Druyvesteyn EEDF. The area above the red line represents the conditions necessary for LTE according to McWhirter's criterion. Reproduced from [135]. (c) Ion acceleration: illustration of the evolution of the dynamic double layer for a plasma plume containing light and heavy ions. Reproduced from [136]. (d) Plasma chemistry: Equilibrium composition of reaction mixture $SiF_4$+$2H_2$ at $P$=1 atm in the temperature range of 300-10000 K. Reproduced from [137]. (e) Plasma radiation: approximate temporal windows for atomic and molecular emission from ns-laser-induced breakdown spectroscopy (top), high-resolution LIBS (configuration: wavelength 1064 nm, energy 35 mJ, beam diameter 9 mm, pulse duration 5 ns) spectra of titanium and polystyrene (bottom). Reproduced from [138, 139].

### *3.1 Plasma expansion*

#### *3.1.1 Hydrodynamic modeling*

At relatively high background gas pressure ($P_{gas}$>$10^2$ Pa), collective processes dominate, and the plasma can be considered as a continuous medium [140]. In this scenario, the fluid dynamics describing a continuous medium are commonly used to depict the expansion evolution of plasma, governed by the principles of mass conservation, momentum conservation and energy conservation. Based on whether the viscous losses and transport phenomena in the flow, as well as thermal conduction, are neglected, these equations are divided into Navier-Stokes equations and Euler equations. Both are frequently employed in simulating expansion processes [141, 142].

In particular, Gornushkin and colleagues assessed the influence of transport phenomena on the evolution of gas ambient plasmas and spectral radiation [134], including viscosity, diffusion between multiple components, and thermal conduction. The results indicated that transport phenomena do indeed affect the mass density distribution within the plasma. However, this seems to be relatively unimportant for transient emission spectra. Similar comparisons have been made in the context of viscosity in liquid ambient atmospheres [143]. Therefore, the Navier-Stokes equations are more appropriate for accurate simulation of viscous effects, thermal processes, and component diffusion. However, the Euler equations can also be effectively utilized when only the evolution of radiation spectra is of concern.

The aforementioned conservation equations are not closed and require additional EOS to solve [144]. The EOS describes the state of a thermodynamic system through the relationship between thermodynamic parameters (including pressure $p$, density $\rho$, temperature $T$, internal energy $\varepsilon$, and entropy $H$) [145]. The ideal gas EOS is the most classical solution that considers the kinetic energy (thermal energy) of atoms or molecules. When considering plasma ionization, the internal energy of the plasma does not completely convert to thermal energy, but rather is “stored” as the ionization potential energy of ionized ions and the excitation potential energy of excited atoms and ions [146]. Specifically, Wen *et al* emphasized its importance for the early evolution of hot plasmas because assuming the plasma as an ideal gas would overestimate its initial temperature, further leading to an overestimation of the corresponding radiation loss and the premature cooling of the plasma. Consideration of ionization potential energy and excitation potential energy allows energy to be stored, enabling the plasma to maintain a relatively high temperature in the later stages of evolution (laser: 1064 nm, 4 ns, 10 mJ, 300 μm) [147].

Interestingly, the use of isothermal or isentropic self-similar solutions to describe the expansion of LIP has been investigated [148-150]. For example, Gornushkin *et al* considered the expansion in vacuum as an adiabatic expansion as a substitute for fluid calculations (laser: 1064 nm, 500 mJ; target: brass; atmosphere: vacuum) [151]. One inconsistency is that the position of the maximum value in the profile obtained from self-similar solutions remains on the target surface. However, Stapleton *et al* measured a series of controlled digital images of laser ablation plumes and indicated that the position of maximum intensity moves along the expansion axis (laser: 266/532/1064 nm, 6 ns, 2-7 J/cm$^{2;}$ target: Li; atmosphere: vacuum) [152]. This may be due to the thermal condition near the target surface is not adiabatic, accompanied by an influx of mass and energy during evaporation. Moreover, the strong recoil pressure from the plume will exert an increasing pressure on the surface. Chen *et al* added the plume volume velocity obtained from the surface pressure term of the Navier-Stokes equation to the isothermal model, based on the conservation of energy and momentum. The simulation results showed good consistency with the temporal and spatially resolved optical experimental diagnostics of the plume [153]. Another issue is the precise determination of the adiabatic index $\gamma$ exhibiting a crucial importance in the self-similarity solution. For low-temperature LIP, a lower value of $\gamma$ comparing with the normal one shows better consistency with experimental results. Marenkov *et al* confirmed the radiation intensities calculated by isothermal and isentropic expansion with $\gamma$=1.25 reproduces the experimental trends compared to the ideal index $\gamma$=5/3 (laser: 1064 nm, 15 ns, 42 J/cm$^2$, 300 μm; target: Li) [154] for monatomic gases. The exact physical explanation behind is still unclear, one possible explanation lies in the ionization/recombination in the plasma energy distribution.

### *3.1.2 Plasma radiation effect*

In the early formation of hot plasmas, the radiation generated by the high temperature due to the injection of laser energy is often significant for plasma expansion [155]. Shabanov *et al* reported that approximately 25% of the initial energy of an ideal gas plasma (including 95% Al, 4% Si, and 1% Mg) with an initial energy of 5 mJ is radiated away after 300 ns (a dynamical model of a laser induced plasma with axial symmetry) [134]. This estimation is made without considering the reheating effect produced by the plasma reabsorbing the emitted radiation. By introducing the radiation heat flux tensor into the energy equation of a pure fluid dynamics model and coupling fluid dynamics unidirectionally to the radiation transport, the radiation loss under given plasma temperature and particle number density is solved. Furthermore, a self-consistent distribution of plasma temperature and local radiation can be provided at the same time step through simultaneously solving the bidirectional coupling of fluid equations and radiation transport equation [148]. Ho *et al* reported approximately 35% of the laser energy absorbed by the plume will be re-radiated into the atmosphere in the form of a continuous spectrum through this method. If this radiation heat transfer is not taken into account, the maximum temperature in the plume region would increase from O(10eV) to O(100eV) (a computer code PLEM (pulsed laser evaporation of metals); laser: 248 nm, 26 ns, 25 J/cm$^2$; target: Al) [156].

When performing specific numerical calculation for the radiation transport equation, the multigroup diffusion approximation is the most practical numerical solution method, which can save computational time and resource compared to a complete analytical treatment. In this framework, absorption coefficients within the specific spectral sub-intervals are assumed to be independent of frequency, and commonly substituted by the Planck mean absorption coefficients. The accuracy of the diffusion approximation solution improves as the number of energy groups increases. Mazhukin *et al* reported that when irradiating an aluminum target with the laser intensity of $10^9$ W/cm$^2$, seven spectral intervals can already achieve the convergence of plasma temperature values in the range of 1.0-200 eV, while higher intensities need to require 61-101 more intervals (the Hartree-Fock-Slater quantum-mechanical model for the photo-ionization cross-sections and the initial spectroscopic characteristics required; laser: 1.06 μm, 10 ns, 200 J/cm$^2$; target: Al vapor) [157]. Radiation losses at different laser wavelengths have also been reported. Mazhukin *et al* observed the highest temperature and the maximum proportion of radiation losses (>30%) in the infrared region at $\lambda$=1.06 μm, with compared to visible region with $\lambda$=0.532 μm and extreme ultraviolet region with $\lambda$=0.248 μm. It was explained by the dominant inverse bremsstrahlung absorption depicting that when the plasma temperature exceeds ~5 eV, its absorption coefficient decreases proportionally to the cube of the optical wavelength (laser: 1.06 μm, 10 ns; target: Al; atmosphere: Ar at low-pressure) [158].

As the plasma gradually cools during its evolution, the impact of radiation on plasma characteristics also diminishes. For typical LIBS plasmas (temperature $10^4$-$10^5$ K, number density ~$10^{19}$ cm$^{-3}$), Shabanov and co-workers estimated the maximum radiation energy loss is less than 0.1% of the plasma's thermal energy by the Stefan–Boltzmann law ($4\sigma T^4/c$) [134]. Marenkov *et al* concluded the absorbed laser energy will redistribute into internal energy and kinetic energy of the plasma in a ratio of approximately 70% and 30%, respectively (multigroup diffusion approximation; laser: 1064 nm, 15 ns, 42 J/cm$^2$, 300 μm; target: Li; atmosphere: vacuum) [155].

### *3.1.3 External magnetic field effect*

Additionally, attention will be focused on the magnetic effect of the interaction between nanosecond pulsed lasers and plasmas, which are often neglected but are frequently discovered through experimental diagnostics instead. Examples include

(i) using external magnetic fields to confine the plasma, further assisting in enhancing laser-induced breakdown spectroscopy (MF-LIBS) [159-161], where the magnetic field is oriented parallel or perpendicular to the axis of plasma expansion [162, 163]; (ii) magnetic-field-assisted laser-induced plasma micro-processing (MC-LIPMM) [164, 165]; (iii) debris reduction using magnetically guided pulsed laser deposition (MG-PLD); (iv) Zeeman effect causing spectral splitting [166]; (v) applying a transverse magnetic field to melt uniformly and refill the ablation cavity with the molten materials [167, 168]; as well as (vi) autogenous magnetic field producing the plume expanding anisotropy, which in turn contributing to polarized emission [169], and etc. The interpretation and modeling of these magnetic effects has also been called for in Bogaerts' review on the mechanism of laser-induced excitation in photochemical analysis [170].

A qualitative explanation based on magnetohydrodynamics (MHD) through two mechanisms: (i) resistive Ohmic/Joule heating, and (ii) adiabatic compression of the plasma by the magnetic field. Specifically, Waheed *et al* explained in detail that the presence of a magnetic field firstly causes ions and electrons in the plasma to be separated by the Lorentz force and further generates an induced current ($\boldsymbol{J}$). The interaction between $\boldsymbol{J}$ and the applied magnetic field $\boldsymbol{B}$ results in the term $\boldsymbol{J}\times\boldsymbol{B}$, slowing down the kinetic velocity and energy of the plasma plume and reheating electrons. This enables to continuously excite ions to higher charge states, the Joule heating effect can be described by the generalized form of Ohm's law in a magnetized ion plasma (laser: 1064 nm, 10 ns, 25-200 mJ; target: $ZrO_2$; atmosphere: vacuum) [171].

Arshad *et al* interpreted the restriction of the plume in the transverse direction of the magnetic field will increase collision rates, leading to further increases in electron number density and plasma temperature, promoting electron collisional ionization and enhancing the ionization fractions (laser: 1064 nm, 10 ns, 25-200 mJ; target: graphite; atmosphere: vacuum) [172]. Hussain *et al* added the plasma beta ($\beta$) to quantitatively describe the scale of the diamagnetic effect for the plasma expansion. One kind of thermal beta ($\beta_t$) is defined as the ratio of particle or thermal pressure ($P_t=n_e k_B T_e$) to magnetic pressure ($P_B=B^2/8\pi$) (laser: 1064 nm, 8 ns, 100 mJ, 10 Hz, 300 μm; target: Al) [173]. When the magnetic pressure equilibrates with the plasma pressure or $\beta_t=1$, the expansion of the plume will cease. Khan *et al* related $\beta_t$ with the plasma expansion velocity, including that lower $\beta_t$ constrains the plume more effectively while higher $\beta_t$ results in ineffective maintenance of the magnetic field, with no constraint in the expansion velocity (laser: 1064 nm, 10 ns, 25-200 mJ, 1.26 mm; target: Co; atmosphere: Ar/Ne) [174].

Furthermore, Rai *et al* proposed a simple emission intensity model based on the expansion velocity and emission duration to explain the enhancement of plasma emission in the presence of a magnetic field (laser: 532 nm, 8 ns, 200 mJ, 10 Hz; target: Al alloy) [175]. Hai *et al* analyzed their experimental results and observed that the emission duration is prolonged and the expansion velocity is reduced under the presence of the magnetic field, leading to an increase in plasma density and emission intensity (laser: 1064 nm, 5 ns, 6.7 J/cm$^2$; target: Al-Li alloy; atmosphere: vacuum) [176].

Harilal *et al* found that the expansion of the plume is not blocked by the magnetic field but is significantly slowed down when $\beta_t$ is approximately equal to 1. It indicated that $\beta_t$ is not the sole determinant to describe the effect of the magnetic field. Additionally, the recoil pressure of the plasma $P_r=nmV^2/2$ is also considered an important factor in describing the magnetic field effect (laser: 1.06 μm, 8 ns; target: Al; atmosphere: vacuum) [163]. Therefore, a directed $\beta_d$ is provided as well to estimate the effect of the magnetic field on plume expansion as the ratio of recoil pressure $P_r$ to the magnetic field pressure $P_B$.

Finally, it should be reminded that although most current explanations are not yet able to accurately and structurally describe magnetic effects and remain widely debated to this day, it is still strongly recommended not to regard these discussions and explanations as exclusive. There are indications that they may have significant contributions to the plasma evolution, and more fundamental research related the magnetic effect of nanosecond pulsed induced plasma is hoped to be inspired.

*3.1.4 Combining particle-based multiscale and hybrid models*

Although more commonly seen in ultra-short pulsed laser modeling, molecular dynamics simulations have also been noted in nanosecond pulse modeling, revealing insights into plasma processes at short time scales on the microscopic level [177-179]. They can provide atomic-level insights into laser-induced processes, offering advantages in simulating specific phenomena such as non-Maxwellian velocity distributions in rarefied gas flows, generation of nano-particles and clusters, and plasma plume splitting [180-183].

On the other hand, Direct Simulation Monte Carlo has also been used to simulate plume expansion dynamics as a mesoscopic particle-based model. In the framework of this approach, the trajectories of specific particles are tracked from the point of view of statistical physics. As an example, combining Direct Simulation Monte Carlo (DSMC) with ionization and radiation absorption models has been employed by both Humphrey and Itina to study the behavior of plasma plume splitting into fast (along the edges of shockwaves) and slow (in the plume core) components for a wide range of background pressure. This behavior can be explained by the fluid-dynamic snowplow effect, resulting a rapidly propagating high-temperature and high-density region between primary and secondary shockwaves [184-186]. Palya *et al* utilized DSMC to investigate the complex structures of shockwaves formed by the interaction between the plasma plume with cavity walls and the re-deposition of ablated materials on the cavity surface under spatial confinement (laser: 266 nm, 10 ns, 22/44.88 μJ; target: Cu; atmosphere: He/Ar/Xe) [187]. Their simulation confirmed that the background gas pressure shows both focusing effect enhancing the efficiency of material removal, and confinement effect restricting vapor condensation within the cavity.

DSMC appears computationally expensive in describing dense plume phenomena, due to the rapid increase in computational memory and time with the increase in the number of simulated particles and grid units [188]. Therefore, Zeifman *et al* proposed a multi-scale model combining computationally cheap fluid dynamics to describe the early dense plume and expensive DSMC simulations for the late-stage plume expansion [189]. Their model can cover the plume expansion from vacuum up to the strong shockwave region under different background gas pressures.

*3.2 Plasma ionization and recombination*

*3.2.1 Population distribution*

The thermodynamic state is crucial as accurate plasma properties and behaviors under different thermal conditions require different accurately thermodynamic computational models to describe them. Complete Thermodynamic Equilibrium (CTE) demands balance between all radiation and collision processes with their reverse counterparts. From different perspectives, it can be categorized into Maxwell equilibrium (for translational velocity and energy), Saha-Boltzmann equilibrium (for atomic and ionic state distributions), and Planck radiation equilibrium (for radiation transport). However, this is an ideal theoretical scenario. For many experimental plasmas such as LIP, which are often not optically thin, causing photon-escaping violating the Planck radiation law. Additionally, laser absorption during the initial formation of LIP may lead to a transient inconsistency in temperature between electrons and heavy particles, which all contravene CTE and at best reach the LTE, otherwise find themselves in collisional radiative non-equilibrium. Wang *et al* used a non-equilibrium model and a local thermodynamic smoothing model respectively to simulate the expansion behavior of laser argon plasma. In figure 5, compared to the LTE model, the plasma expansion rate in the non-equilibrium model is faster and the plume temperature is higher [20].

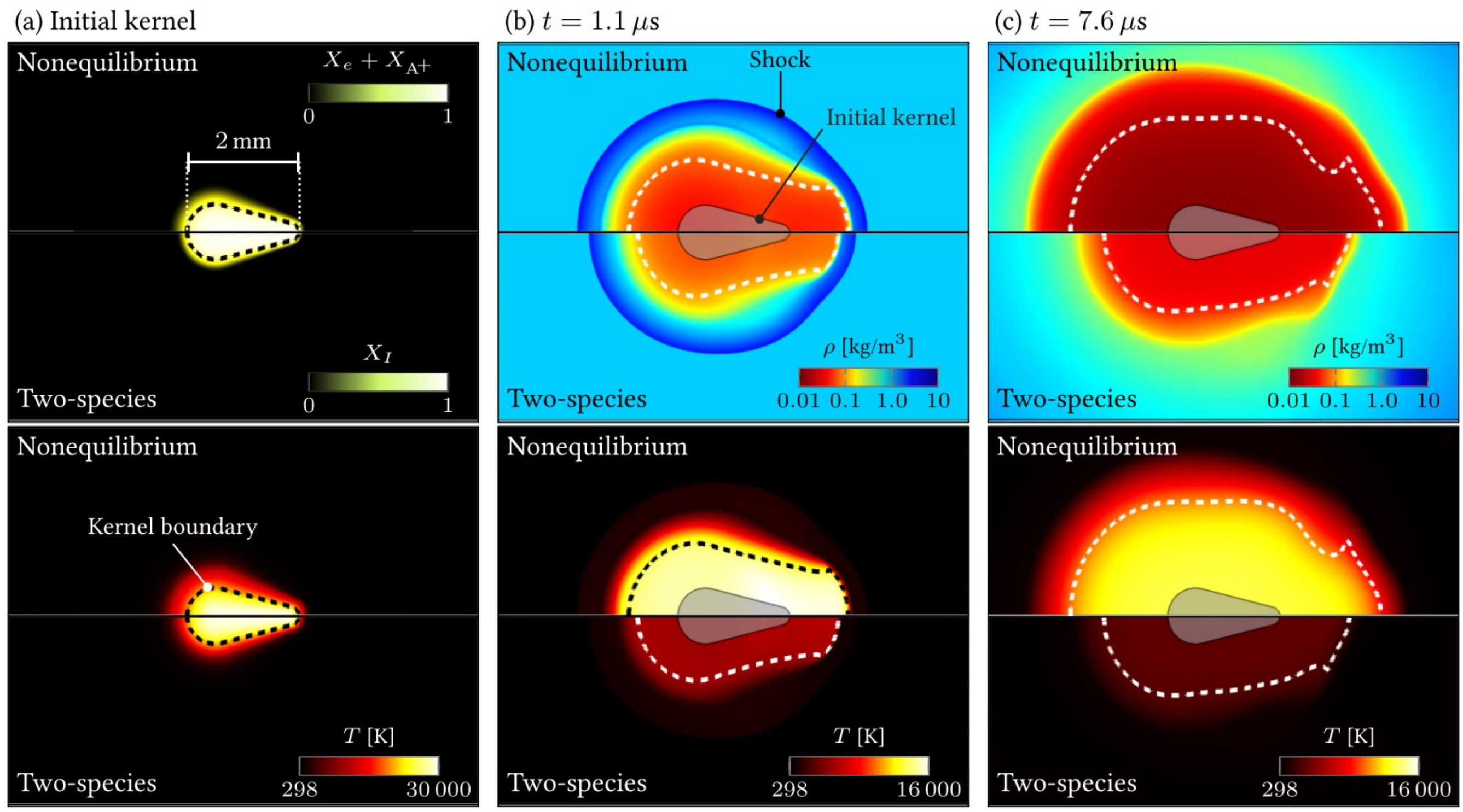

**Figure 5.** The expansion process of argon plasma in 0.5 atm at (a) its initial state, (b) $t$ = 1.1 μs, and (c) $t$ = 7.6 μs, computed using the non-equilibrium and two-species (LTE) models as labeled. The laser-induced argon plasma size is 2 mm and the initial temperature is 30000 K. Dashed lines denote the kernel boundary. Note the change in temperature scale. Reproduced with permission from [20].

1) Local Thermodynamic Equilibrium

If momentum and energy transfer between plasma species are dominated by electron collisions, it can be considered in LTE. Many studies consider the minimal electron number density as the boundary of electron collision dominance. If the electron density in the system exceeds this critical value, it can be considered that the energy loss due to radiative escape is far less than the energy exchanged in collision processes. Therefore, electron collision is the main energy transfer way in plasma, and the assumption of LTE status in plasma is valid. This critical value which is derived by Griem [190] implies that collisional excitation/de-excitation processes are at least ten times faster than radiative processes, also known as the McWhirter criterion, expressed as:

$$n_e > 1.6\times10^{12}\sqrt{T_e}\cdot\Delta E_{\mathrm{lu}}^3 \tag{1}$$

where $n_e$ (cm$^{-3}$) and $T_e$ (K) are the electron number density and electron temperature, respectively, and $\Delta E_{\mathrm{lu}}$ is the maximum potential energy gap between adjacent levels, typically between the ground state and the first excited state, i.e., the Lyman transition $\Delta E_{2\text{-}1}$ expressed in electron volts. This formula is only applicable for optically thin plasmas, where self-absorption effects can be ignored.

The electron number density plays a crucial role in determining plasma properties, but assessing LTE solely through the above simplified description is inadequate. Due to LIP expanding followed by energy exchange with ambient gas, its transient nature and spatial gradients may disrupt equilibrium. Cristoforetti *et al* [191] have modified the McWhirter criterion using thermodynamic relaxation time $\tau$ and diffusion length $\Lambda$. If the relaxation time for re-establishing Saha-Boltzmann equilibrium is much shorter than the thermodynamic evolution time of LIP, the plasma can still be considered quasi-steady.

The condition is expressed as:

$$\left|\frac{T_{\mathrm{e}}(r,t+\tau)-T_{\mathrm{e}}(r,t)}{T_{\mathrm{e}}(r,t)}\right|<<1,\ \left|\frac{n_{\mathrm{e}}(r,t+\tau)-n_{\mathrm{e}}(r,t)}{n_{\mathrm{e}}(r,t)}\right|<<1 \tag{2}$$

where $\tau$ is the time needed for the establishment of excitation and ionization equilibria, $r$ and $t$ represent the spatial position coordinates and system time, respectively.

Meanwhile, the diffusion length of atoms and ions in the plasma relaxation time must be much smaller than the length of thermodynamic property changes. The thermodynamic non-equilibrium caused by strong particle density gradients is described mathematically as:

$$\left|\frac{T_{\mathrm{e}}(r+\Lambda,t)-T_{\mathrm{e}}(r,t)}{T_{\mathrm{e}}(r,t)}\right|<<1,\ \left|\frac{n_{\mathrm{e}}(r+\Lambda,t)-n_{\mathrm{e}}(r,t)}{n_{\mathrm{e}}(r,t)}\right|<<1 \tag{3}$$

where $\Lambda=(D\cdot\tau)^{1/2}$ is the diffusion length during the relaxation time $\tau$. $D$ is the diffusion coefficient of LIBS plasma.

The McWhirter criterion can be quickly tested with plasma temperature and electron number density, but equations (2) and (3) are much more difficult to test experimentally because they require information about the temporal evolution and spatial distribution of plasma parameters [192].

For electron collision-dominated LIP, if equilibrium is established faster than thermodynamic parameter evolution, it can be determined as LTE or quasi-LTE. At this point, energy loss due to radiation is less than that involved in other processes of substance types, making Saha-Boltzmann and Maxwell distributions still effective descriptions of atomic state population distribution in LIP. Here, it should be noted that during the LIP evolution (maybe lasting a few microseconds), both electron number density and excitation temperature will decrease, and transitions of some energy levels will depart from collisional equilibrium. At this stage, the electron number density of the system is lower than the critical value calculated for certain energy levels using equation (1). Therefore, the equilibrium relationships corresponding to these energy levels no longer apply. The breakdown of equilibrium makes the plasma not in LTE state, but rather in partial local thermodynamic equilibrium (pLTE). In such cases, collisional radiative models may be required to study the population distribution of the plasma.

2) The collisional radiative non-equilibrium

In recent years, there has been a noticeable trend towards collisional-radiative modeling within LIP focusing on non-equilibrium processes. Unlike LTE, in collisional radiative condition, the population distributions of various states do not follow the Boltzmann distribution function. In contrast, the density of a given excited particle should be considered by solving a set of rate equations that account for all atomic processes that increase or decrease the particle abundance [193]. Before delving into the specific solution of rate equations, it is important to revisit twelve atomic processes considered significant in laser-plasma interactions to understand the conditions under which collisional radiative processes occur. Each direct process has its corresponding reverse process, such that the initial states of interacting particles are the final states of the reverse process, and vice versa. There is a total of six pairs of combinations, which can be categorized into collisional and radiative interactions, as shown in table 1. Detailed explanations of each process can be found in [194].

**Table 1.** The types of collision reactions in laser-induced plasma. Reproduced from [194].

| Reaction | Direct/inverse process | Type |
|---|---|---|
| $A_m^{+\zeta}+e\Leftrightarrow A_m^{+\zeta+1}+e+e$ | Electron impact ionization/3-Body recombination | C |
| $A_m^{+\zeta} + e \Leftrightarrow A_{m'}^{+\zeta} + e$ | Electron impact excitation/Electron impact deexcitation | C |
| $A_{m'}^{+\zeta} \Leftrightarrow A_m^{+\zeta} + hv$ | Spontaneous decay/Resonant photoabsorption | R |
| $A_m^{+\zeta} + hv \Leftrightarrow A_{m'}^{+\zeta+1} + e$ | Photoionization/Radiative recombination | R |
| $A_{mm'}^{+\zeta} \Leftrightarrow A_0^{+\zeta+1} + e$ | Autoionization/Dielectronic recombination | R |
| $A_m^{+\zeta} + e \Leftrightarrow A_m^{+\zeta} + hv$ | Bremsstrahlung/Inverse Bremsstrahlung | R |

Here C means 'collisional' and R means 'radiative'.

When the collisional (C-type) processes reach equilibrium far ahead of the radiative (R-type) processes, the population of atoms and ions in various states is entirely controlled by electron collisions. This collision-dominated plasma states in the LTE as introduced in the previous section. However, when the radiative processes of R-type cannot be neglected compared to the collisional processes of C-type, especially in the early stages of laser pulse irradiation when radiation energy occupies a significant portion of the plasma energy, a comprehensive description of collisional and radiative processes is required, necessitating a collisional-radiative model [195]. Furthermore, Capitelli *et al* classified specific collisional radiative processes into three scenarios: steady-state (CR-SS), quasi-steady-state (CR-QSS), and time-dependent (CR-TD), based on the evolution over time of population and rate of population reduction of different excited state particles [193].

The difficulty in solving rate equations lies in obtaining the accurate rate coefficients, which requires establishing internal level distributions for each species, as well as electron collision cross-sections and radiation transition probabilities [196-198]. This process is generally too complex for a straightforward solution, particularly for medium *Z* to high *Z* elements, and is continually being refined and improved. Rate coefficients are typically obtained from two sources: experimental diagnostics and model calculations. Rui *et al* demonstrated ion storage rings and electron beam ion traps (EBIT) or electron beam ion sources (EBIS) as ideal experimental environments and performed studies on electron-ion collision cross sections spanning from meV to several keV [199].

Chung *et al* utilized detailed balance principle (DBP) to derive reverse process rate coefficients from direct process rate coefficients [200]. The Multiconfiguration Dirac-Hartree-Fock (MCDHF) method [201-203], Quantum Electrodynamics (QED) [204], Relativistic Many-Body Perturbation Theory (MBPT) [205], Screening Hydrogenic (SH) approximation [195], and Relativistic Distorted Wave (RDW) method [199, 206] offer theoretical solutions for obtaining ion energy levels and photoionization cross-sections with spectral precision. For example, Srivastava *et al* utilized the relativistic distorted wave (RDW) theory to calculate electron collision fine structure ionization cross sections for medium Z elements such as Zn, Mg, Mo, Si, Cu, and by solving quasi-steady-state collisional radiative equation, population distributions are obtained and subsequently line emission intensities matches spectral measurement values [202, 206-209]. By combining rate equations with fluid dynamics or molecular dynamics simulations describing plasma expansion as outlined in Section 3.1, the complete evolution of non-equilibrium plasma expansion, collision, and radiation processes can be fully described. Gornushkin [210] and Cristoforetti [191, 211] also provided clear explanations of collisional-radiative models in their reviews.

Additionally, under non-LTE conditions, the electron energy distribution function no longer follows a Maxwell-Boltzmann distribution. If one still derives electron temperature $T_e$ or number density $n_e$ based on the Saha equation assumption from spectra, there may result in deviations in experimental results. Liu *et al* utilized the Druyvesteyn EEDF to simulate non-LTE plasmas, showing better fitting with experimental line profile compared to a Maxwellian EEDF (laser: 1064 nm, 9-20 ns, 20-72 mJ; target: Si; atmosphere: vacuum) & (laser: 532/1064 nm, 7.2/8 ns, 68/79 mJ; target: Si) [135, 148]. In their investigation, deviations from LTE trends were confirmed by calculating local $T_e$ and $n_e$ using the lines of Si III and Si IV through the McWhirter criterion.

*3.2.2 Ion acceleration and space charge effect*

It is worth mentioning that in Section 3.1 describing the fluid dynamics model of plasma expansion and molecular dynamics simulations, regardless of whether plasma radiation or external magnetic fields are considered, the various components within the plasma are assumed to have the same directional velocity of motion. However, it has been widely reported that there is a common phenomenon of ion acceleration in multi-component plasma flows, including two typical effects: mass effect and space charge effect.

(i) Mass effect: Lighter ions are scattered to the edge of the plasma flow, while heavier ions form the core of the plasma. Ojeda-GP *et al* reported a direct positive relationship between ion mass and kinetic energy after studying five multicomponent materials (laser: 248 nm, 2 ns, 2/3 J/cm$^2$; target: ceramic; atmosphere: vacuum/0.01 mbar $O_2$/0.1 mbar $O_2$) [212]. Ahmad *et al* investigated the angular distribution anisotropy of this acceleration effect(laser: 532 nm, 6 ns, 4-7.1 GW/cm$^2$, 900 μm; target: Si/Ge; atmosphere: ultrahigh vacuum) [213]. Yao *et al* analyzed the ion energy distribution (IDE) measurements and found that heavy ions exhibit a high-energy tail while light ions show a high-energy peak. Moreover, as the angle deviating from the normal direction of laser incidence increases, ion energy gradually decreases, the high-energy tail of IDE for medium-mass ions narrows and transforms into a high-energy peak distribution (laser: 248/308 nm, 20 ns; target: Cu/AuCu/$La_{0.33}Ca_{0.67}MnO_3$/$LiMn_2O_4$; atmosphere: vacuum) [136].

(ii) Space charge effect or charge effect [214]: The higher the ion charge state, the faster the expansion velocity of ions, and the farther their spatial distribution extends. There is also a strong positive correlation between the average ion energy and the ion charge state in terms of IDE. As the charge state increases, the ion energy detected by the ion energy analyzer (ICE) continues to increase, the ion yield collected by the ion collector (ICs) gradually decreases, and the appearance in the time-of-flight (TOF) detection occurs earlier [215-218].

On the one hand, such particles segregation with different masses and charge states poses challenges for processes like PLD, which is a powerful and versatile method for producing multi-component material thin films, nanoparticles, and clusters suitable for various applications. This is possible due to the fact that the heterogeneity in the plasma flow makes the process of stoichiometric deposition uncontrollable, leading to defects in thin film growth, and more issues [219, 220]. On the other hand, it also poses a risk to the precise calibration of LIBS. Because this analytical atomic emission spectroscopy strongly depends on the spatial distribution of elements in the transient plasma. Obviously, the particle separation phenomenon affects the spatial distribution of elements [221].

In the following discussion, several promising mechanisms are summarized to explain the ion acceleration within the plasma induced by pulsed lasers for medium laser intensity (~ $10^9$ W/cm$^2$) and nanosecond (~ ns) pulses. At the same time, we also recommend the review of Pirozhkov [222] which detailed many established and potential mechanisms for ultra-short, high-intensity laser-driven proton/ion acceleration, contributing to understanding the fundamental physics processes of ion acceleration. It has also been considered as one of the fruitful areas in the field of the laser-target-plasma interaction thanks to the invention of chirped-pulse amplification (CPA) technology and plasma-based ion accelerators to explore high-energy density physics, even though the laser intensity (>$10^{17}$ W/cm$^2$) and timescale (<10 ps) are not fully consistent with the range of our focus on this review.

1) Shifted Maxwellian-Coulomb distribution

This explanation suggests that the internal electric Coulomb field in the plasma accelerates ions, resulting in an ion energy distribution that is a combined effect of plasma temperature and internal Coulomb interactions, Torrisi *et al* called this relation "the shifted Maxwell-Coulomb distribution" [223]. In this case, the ion velocity is a combination of flow velocity and Coulomb acceleration (laser: 1064 nm, 9 ns, 900 mJ, 30 Hz; target: Ta; atmosphere: vacuum) [224, 225]. On the contrary, the energy distribution of neutral particles depends on the Maxwell distribution, determined only by the plasma temperature.

2) Double layer effect

In the previous discussion, the shifted Maxwell-Coulomb distribution posits that the increase in ion velocity is attributed to the Coulomb field generated within the plasma, the researches of Skočić *et al* and Wang *et al* provide an explanation for the Coulomb-field generation as a double layer on the order of the Debye length resulting at the plasma edge for the charge separation (laser: 532 nm, 5 ns, 125-400 mJ, 1 Hz; target: Cu; atmosphere: air/Ar) [218, 224]. Specifically speaking, high-energy electrons within the plasma possess thermal velocities faster than ions, once the laser energy being absorbed into the initial plasma as an energy reservoir (alternatively, it is suggested that both hot and cold electrons simultaneously exists in the plasma), electrons will be the first to escape to the quasi-neutral front, thus forming an electron-rich outer layer and an ion-rich inner layer at the boundary.

Yao *et al* derive qualitatively the velocity gain $\Delta v$ that ions obtain when crossing the double layer by applying Newton's second law [136] ,

$$m\Delta v = qE\Delta t \tag{4}$$

where $m$ is the ion mass, $q$ is the ion charge, $E$ is the electric field intensity, and $\Delta t$ is the time ions spend within the double layer.

As for the mass effect interpretation, when assuming that light and heavy ions have a similar absorption effect on laser energy, the initial energies of light and heavy ions are the same at the early stage. Therefore, the lighter ions will acquire a faster initial thermal velocity, denoted by subscript $l$, compared to the heavier particles denoted by subscript $h$. Alternatively, $v_l/v_h$ is proportional to $(m_h/m_l)^{0.5}$, indicating that when the charge states are equal, only the ions with different masses will have a ratio approximately $(m_l/m_h)^{0.5}$ of the time spent in the electric field within the double layer. As a result, lighter ions spend less time in the electric field, making them more easily enriched at the edge of the plasma, while less Coulomb potential energy will be converted into kinetic energies of lighter ions. However, heavier particles with smaller thermal velocity will spend more time within the double layer, making up more positions in the plasma core and obtaining a greater kinetic energy gain.

Similarly, in terms of the charge effect, ions with higher charge states experience a greater velocity gain, manifesting in space-resolved spectra or time-of-flight measurement as faster motion of ions in higher ionization states. It is worth mentioning that the double-layer effect has been observed more pronounced under ultra-short pulses and higher laser energies [214, 226], well-known as the Target Normal Sheath Acceleration (TNSA) model. The plasma edge forming the double layer is also referred to as the Transient Dynamic Sheath in the review of Daido [222].

3) Particle-in-Cell (PIC) Simulation

The above concept of double layer has been further visualized by 1D particle-in-cell simulation to describe ion acceleration in a nanosecond domain and $10^{12}$ W/cm$^2$ of power intensity range. In the research of Mascali *et al*, they first observe 1) prompt electrons emission, 2) plasma plume fragmentation, and 3) the presence of a two-electron-temperature (TET) inside the plasma using a movable Langmuir Probe in experiments (1064 nm, 600 mJ, $10^{12}$ W/cm$^2$, 6 ns, Al, vacuum (4 $\times$ $10^{-6}$ mbar)). The PIC method based on the mean field approximation over a 1D lattice is used to reproduce all features of the experimental results described above. In each simulation, the initial quasi-neutrality is maintained only globally in the lattice and not locally. The initial velocity distributions of electrons and ions are shifted Maxwellian velocity distributions.

Their PIC simulations show that the "*ab initio*" TET plasma hypothesis (accompanied by no initial spatial displacement between electrons and ions) cannot produce a prompt electron bunching or plume fragmentation. In contrast, PIC simulation with only one initial electron temperature and the presence of an electron layer at the plasma front (i.e., accompanied by an initial spatial ion-electron displacement) has been observed with all the experimental features. Firstly, the prompt electron escapes rapidly and decouples from the rest of the plasma. Thereafter a superthermal electron component appears within the plasma bulk. Two different components of the electron velocity distributions (hot and cold) correspond to a TET plasma. Complex plume fragmentation and nonlinear ion acceleration are consequently activated [227].

4) Multi-fractal fluids

Another different explanation is that the uneven spatial distribution of light and heavy elements is caused by particle collisions. At the same time, particle collisions cause energy to transfer from the lighter plume to the heavier one [228]. Irimiciuc *et al* simulated a series of particle collision and energy transfer phenomena [229-232] in their studies based on the moving multifractal paradigm combined with hydrodynamic equations. In their method, the physical variables used to describe plasma dynamics are functions of spatial-temporal coordinates and scale resolution. Particles at different scale resolutions exist simultaneously, and each of them moves along its continuous and non-differentiable curve in the multi-fractal spatial domain. Specifically speaking, this scale resolution is closely related to the number of collisions particles undergo during the expansion process. Higher scale resolution corresponds to higher collision frequency, increasing particle thermal velocity and widening distribution range. Thus, one can simulate element segregation in multi-component plumes due to ion acceleration effects by extracting different fractal dimensions for light and heavy ions with different charge states.

### *3.2.3 Plasma chemistry*

With the expansion of the nanosecond LIP plume and energy exchange with the surrounding gas, the temperature decreases, inducing recombination and reducing the number of charged particles in the plasma domain. When the temperature drops sufficiently, collisions between heavy particles (atoms and ions) become significant compared to electron collisions, leading to the formation of molecular species. In recent years, studies on molecule formation [233] in organic laser-induced plasmas and halogen sources in inorganic materials [234] have received much attention due to the development of LIBS in detecting molecular species. Therefore, there is an expectation to simulate the particle evolution during LIP expansion and the process of molecular spectrum formation through chemical reaction modeling. It can promoting experimental studies of molecules, radicals and ions in the nanosecond laser-target interactions.

1) Equilibrium chemical reaction modeling

Chemical equilibrium conditions are more suitable for describing the LIP plume typical under LIBS conditions, generated by nanosecond lasers in air or inert gases at standard atmospheric pressure. Under the LTE assumption, where chemical reaction rate coefficients are assumed infinitely large and reach chemical equilibrium instantaneously, ionization-recombination and excitation-deexcitation in the plasma follow the Saha-Boltzmann equilibrium. This framework can describe the expansion of single-component metallic LIP in inert gases or vacuum. When LIP expands in reactive gases or when the target itself is gas-phase molecules, chemical dissociation and recombination equilibrium of molecules need to be considered. Gornushkin and Shabanov extended the Saha relationship to plasma equilibrium composed of multiple species ($N_2$, $C_2$, $Si_2$, CN, SiN, SiC molecules, and other atomic ions) based on the Guldberg-Waage law, utilizing *ab initio* methodology for molecular partition functions [235]. They further extended the Saha equation to equilibrium formation of negative ions and found that negative ions and molecular ions have minimal impact on LIP mass and energy density. Solving chemical equilibrium for multi-component LIP chemical systems is challenging due to the large temperature variations and expansion spatial scales. Shabanov *et al* developed an equation of state solver for equilibrium chemical reaction models, simulating LIP of the C-N-Si ternary system in the temperature range of 300-15000 K [236].

Combining with fluid dynamics codes, equilibrium chemical reaction models provide information on particle distribution in space and the rules governing particle number density with temperature changes. Dors *et al* coupled $N_2$ chemical dissociation under LTE assumption with fluid dynamics to study the flow phenomena of laser-induced sparks in nitrogen atmospheres (laser: 1.064 µm, 5 ns, 100 GW/cm$^2$; target: $N_2$) [237]. Hermann *et al* utilized this model to study the expansion of multi-component plasma plumes of Fe, Si, Al, Mg in $O_2$ environments (laser: 266 nm, 4 ns, 40 mJ, 150 mm; target: Al; atmosphere: Ar) [140]. They found that the chemical reactions forming AlO radicals only occur at the outer edge of the inhomogeneous plasma in air. This emphasizes the importance of chemical reactions in the calibration-free analysis of organic materials. Veiko *et al* applied this model to simulate laser ablation of titanium in air and combined experimental findings to discover that condensation of metal oxide onto the ablation surface from the plasma is a crucial mechanism for forming nano-porous oxide films on metal surfaces (laser: 1064 nm, 100 ns, 58-300 mJ, 60 kHz; target: Ti) [90]. However, the described chemical reaction models are decoupled from the initial formation of the LIP plume, assuming that the timescale for multiphoton ionization is much shorter than that for electron collision ionization, setting the initial shape of the plasma and its temperature and density directly from experimental data. Predicting the full effects of material ablation and multiphoton ionization will still be challenging.

2) Non-equilibrium chemical reaction modeling

When LIP does not satisfy the equilibrium assumption, non-equilibrium chemical kinetics models are adopted to describe particle evolution during LIP expansion with a set of chemical reaction systems. This system includes general reactions within LIP and their corresponding rate coefficients, which are finite and temperature-dependent. Depending on the types of chemical reactions, determination of their rate coefficients involves different quantum chemical theories. For a given reaction (non-electron collision), theoretical calculations are required based on the reaction types and various quantum chemistry theories. Additionally, Catoire *et al* conducted high-temperature reaction experiments in a high-temperature reaction furnace, fitting chemical equations to obtain the Al/$O_2$ chemical reaction system through appropriate analytical techniques [238]. Casavola *et al* calculated the complete Ti/$N_2$ reaction system using corresponding quantum chemical computational theories [130]. Finko *et al* established early-stage chemical kinetics models of U/O evolution, addressing the diagnostic challenges of U oxide in LIBS and aiming to improve U detection accuracy and isotopic detection (nuclear medical applications) in LIBS [239]. Their model was based on metal-oxygen reaction systems such as in [238] for Al/O reaction systems, and combined with the simple collision theory (SCT) to invert the U/O reaction system. The method of combining the chemical reaction system with a fluid dynamics model is detailed in [240]. It allows for the simulation of species evolution as laser-ablated U plasma propagates in the air atmosphere. As shown in figure 6, the simulation results indicate that the cool environment at the plume outer edge is conducive to the formation of $UO_3$. Conversely, the high-temperature environment in the core of the plume causes UO and $UO_2$ to concentrate in the core, forming an overlapping layer of oxides. This simulation prediction has also been confirmed in [241].

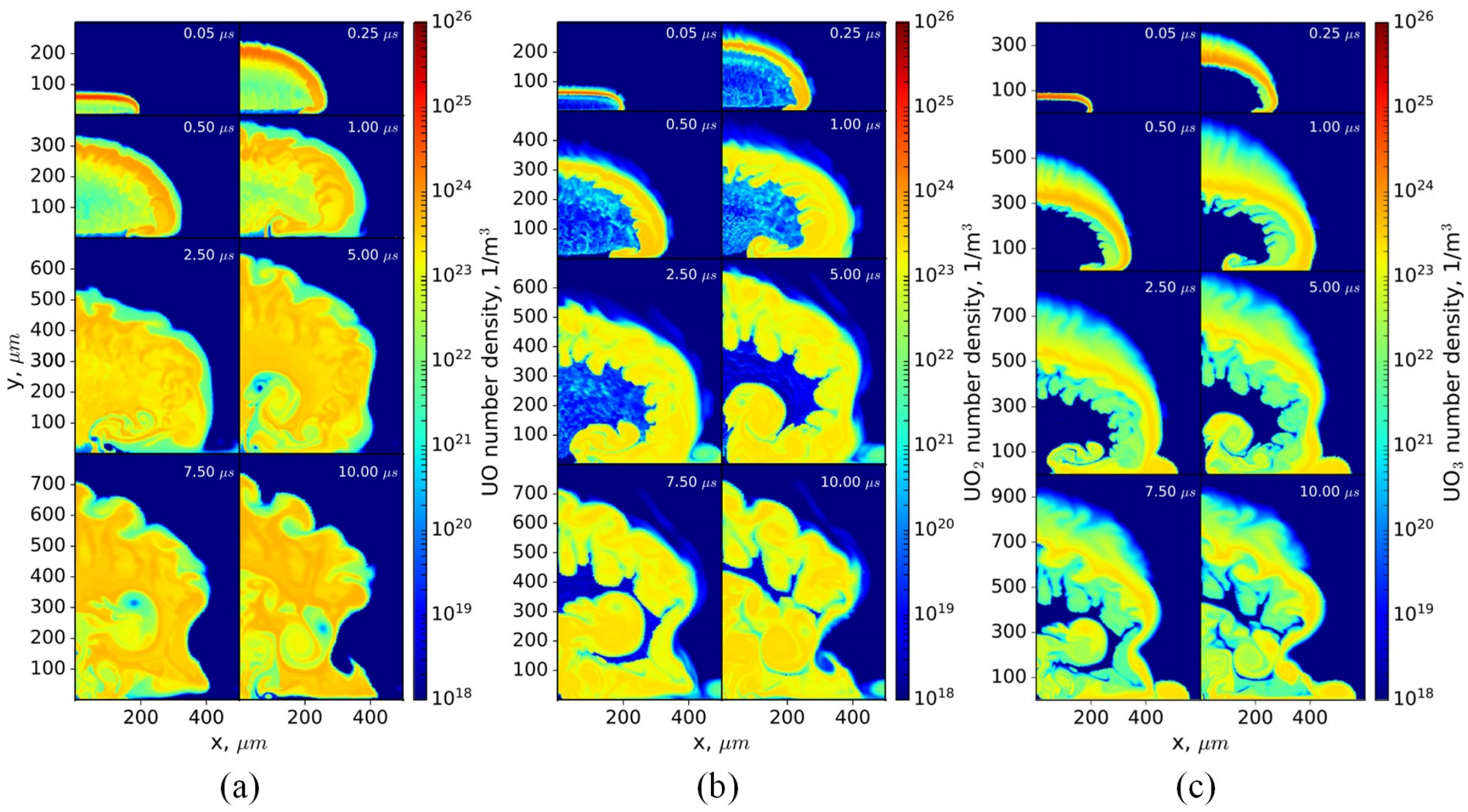

(a) (b) (c)

**Figure 6.** Snapshots of (a) UO, (b) $UO_2$, (c) $UO_3$ number density profiles during the first 10 μs of plume expansion. The shape of the initial plume is an oval with a radius of 170 μm and a depth of 5 μm. The plume is uniformly filled with 35 MPa, 1 eV atomic uranium, while the rest of the domain is uniformly filled with 1 atm, 300 K $O_2$. (Reproduced from [240]).

### *3.3 Plasma radiation*

#### *3.3.1 Atomic Emission*

The atomic emission spectroscopy consists of two parts: continuous spectrum and line spectrum. Line spectrum is widely used in LIBS element composition analysis, while the continuous spectrum is important for fully ionized plasma. Additionally, due to the continuous presence in time and sensitivity to plasma temperature, continuous spectrum is increasingly considered in plasma diagnostics [242].

The emission rate of the line spectrum for optically thin plasma (where the plasma size is negligible compared to the mean free path of radiation photons) is closely related to the ion number density of the initial upper ionization state, which is determined by the population distribution of the plasma. While the population distribution of the plasma is controlled by the Boltzmann distribution under LTE and obtained by the closed solution of rate equations under collision-radiative equilibrium [195]. The details have been explained in the previous section 3.2.1 on the ionization process.

However, when considering the non-negligible optical thickness of the plasma, emitted photons may undergo absorption and re-emission. In such cases, the solution of the radiation transport equation will be necessitated [243] . Burger *et al* utilized the analytical solution of the one-dimensional form in such a defined spatial domain, where the plasma hemisphere is divided into two regions: the high-temperature hot core and the low-temperature cold periphery (laser: 1064 nm, 6 ns, 75 mJ; target: U; atmosphere: $N_2$) [117]. This partitioning provides a first-order correction for line spectrum calculations compared to a uniform plasma. Furthermore, D'Angelo *et al* utilized a more real non-uniform plasma with finer spatial partitioning and temperature distribution approximated by a second-order parabola to reconstruct the self-reversal of Co I 340.51 nm emission line (laser: 1064 nm, 7 ns, 35-75 mJ, 2 Hz; target: Co-Cr-Mo alloy) [244].

On the one hand, the evolution of spectral line profiles and optical thickness can be computed by providing the plasma temperature and electron density (or pressure). For instance, in figure 7 (c), Burger *et al* synthesized spectra for U I and U II transitions through comparing with the measured spectra. The aim is to determine the Stark broadening parameters for high-Z elements [117]. Hermann *et al* calculated the line profiles and optical thickness of low-temperature aluminium atomic spectral lines (<8000 K) (simplified analytical solutions of the radiation transport equation; laser: 266 nm, 4 ns, 40 mJ, 80 J/cm$^2$, 100 μm; target: Ti-sapphire-Al alloy; atmosphere: vacuum) [245]. They elucidated the difficulties in diagnosing plasma temperature and electron number density using resonance lines. Specifically, due to the optical thickness of resonance line being much larger than that of transitions in highly excited states, the saturation plateau of self-absorbed resonance lines will approach blackbody radiation brightness earlier. This makes plasma temperature diagnostics based on line intensity ratios challenging. Both electron density and Stark broadening increase gradually with temperature. However, the width of resonance lines decreases sharply due to the reduction in neutral atom densities. The two factors obscure the slight increase in Stark broadening. Therefore, resonance lines are also unsuitable for electron density measurements based on the linear relationship between Stark broadening and electron density.

On the other hand, the early evolution behavior of high-energy electrons and ions in LIP remains a focus of fundamental research, the use of synthetic spectra to infer experimental plasma temperature and electron density is considered convenient, cost-effective, and efficient. In figure 7 (a) and (b), Burger *et al* resolved Balmer spectral lines of hydrogen and deuterium and assessed the deviation from the equilibrium state through fitting the experimentally measured spectra with the synthetic spectra (laser: 1064 nm, 10 ns, 140 mJ, 10 Hz, 0.3 mm; target: $D_2O$ with deionized light water; atmosphere: air) [246].

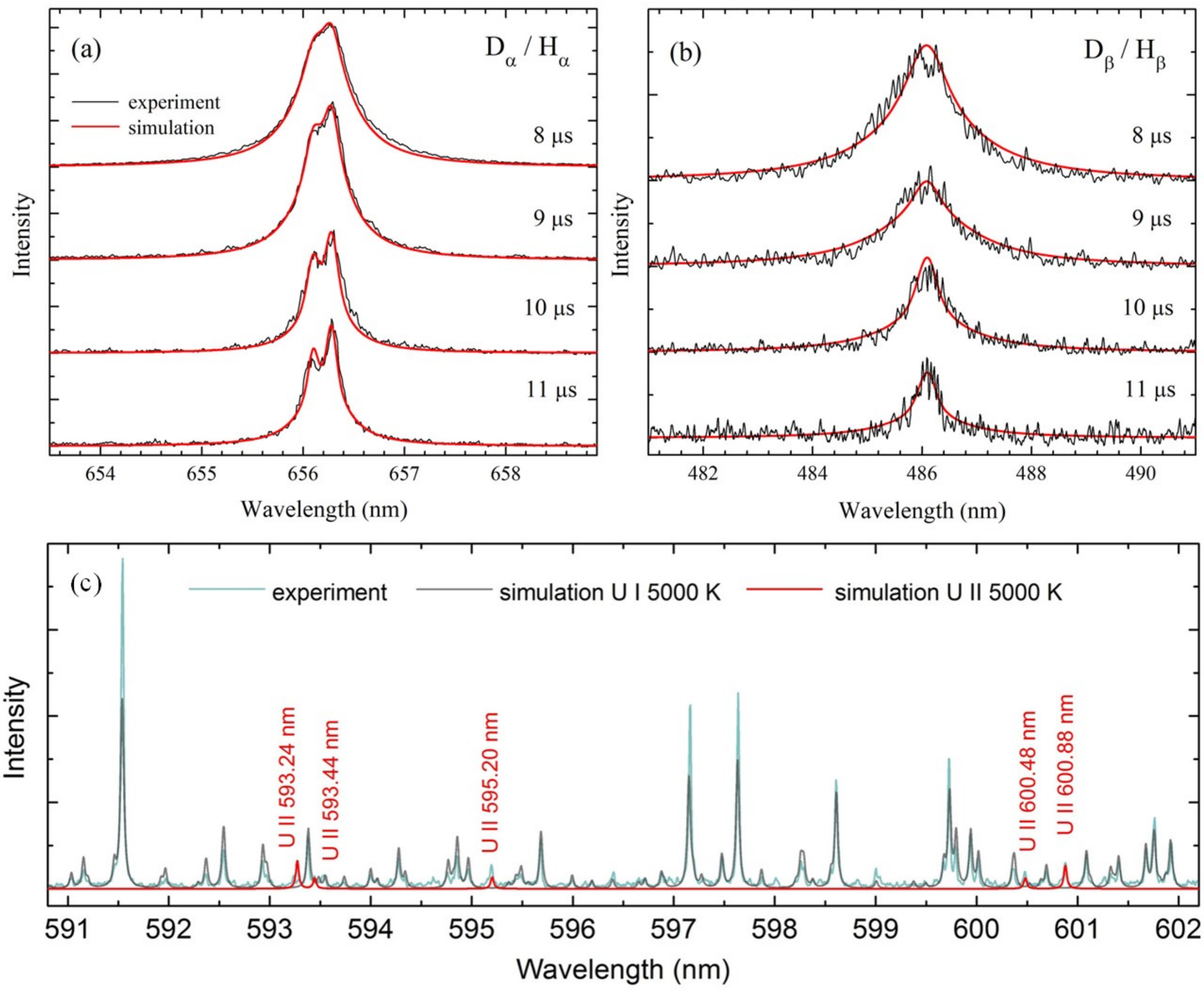

**Figure 7.** Measured and computed spectral line profiles of hydrogen deuterium plasma for Balmer (a) alpha and (b) beta spectral lines at various delays after the laser pulse (1064 nm, 10 ns, 75 mJ). (Reproduced from [246]). (c) Generated uranium plasma spectrum for 5000 K and experimental (1064 nm, 6 ns, 75 mJ, 1 mm) spectrum recorded at 1000 ns and 2 mm away from the target (Reproduced from [117]).

*3.3.2 Molecular emission*

In the core region of LIP, high-temperature ablated species predominantly exist in neutral and ionized states. However, the plasma temperature distribution is not uniform, with cooler regions existing at the periphery of the plume, allowing various complex molecular species to survive [247]. As the plasma gradually cools over time, atomic recombination leads to further promotion of molecular formation, with molecular emissions characterized by vibrational and rotational transitions becoming prominent [248]. This aids in the correct understanding of the formation process of molecules within the cooling plasma and the information regarding chemical bonds between atoms. The spectral region in mid- to long-wave infrared (2-12 μm) spectra is referred to as the fingerprint region characterized by vibrational-rotational transitions. This region is highly sensitive to slight differences in molecular structure, preserving direct information about the composition of molecules compared to the cluttered background of ultraviolet/visible/near-infrared spectra [249, 250].

Molecular spectra are also commonly employed to assess the equilibrium state in transient plumes. Specifically, Parigger *et al* evaluated the thermal and chemical equilibrium states of the plasma, through comparing and fitting the measured results of molecular spectra with numerical simulation results to estimate plasma vibrational and rotational temperatures, as well as the molar fraction ratios of various molecular species (laser: 1064 nm, 14 ns, 190 mJ, 10 Hz) [251]. Numerical simulations often involve solving the radiative equation for a single zone of continuous medium temperature, disregarding optical thickness and radiation reabsorption. Additionally, Ankerhold *et al* introduced a reactive force field (ReaxFF) method based on bond-order calculations, which provides atomic-scale molecular dynamics explanation for CaO molecules formation (laser: 1064 nm, 2.5 ns, 3 mJ, 40 Hz; target: CaO) [252].

Regarding thermal equilibrium estimation, Chemin *et al* confirmed the non-equilibrium energy distribution in rotational and vibrational degrees of freedom in AlO molecules (laser: 266 nm, 5 ns, 30 mJ, 20 Hz; target: mono-crystal of Al) [253]. In their research, the calculated fluorescence spectrum is synthesized by directly measuring the rotational temperature of the ground electronic state $X^2\Sigma^+$ through fluorescence, and the rotational and vibrational temperatures of the excited state $B^2\Sigma^+$ through laser collision-induced fluorescence measurements. This is achieved by nonlinear least squares fitting to match the experimental fluorescence spectrum.

For chemical equilibrium, Harilal *et al* observed the significant discontinuities in the estimated temperatures of $CN/N_2^+$ and OH bands around ~30 μs during the lifetime of laser-induced air plasma, which is considered to violate LTE conditions (laser: 1064 nm, 6 ns, 55 mJ, 14 μm; target: air) [254]. Meanwhile, deviations up to 2-6 times from calculated values were observed in the measured mole fraction ratios of $N_2^+/N_2$ and $CN/N_2$ at higher temperatures (6000-10000 K) at the same time. Based on thermal and chemical equilibrium, this deviation is attributed to the plasma cooling faster than the time required for chemical reactions to adjust the gas-phase composition to maintain chemical equilibrium.

Molecular emission is also used to determine the isotopic abundance ratios in samples, this new technique is termed laser ablation molecular isotopic spectroscopy. Russo and his colleagues conducted a series of studies on this. As an overview, in carbon isotope measurements, simulation of the swan system of $C_2$ within the ranges of 473.5-476.5 nm and 466-478 nm (including the (1-0) band of the isotopologue $^{13}C^{12}C$) yielded a $^{13}C/^{12}C$ ratio (1.08%) that closely matched the value measured by isotope ratio mass spectrometry (1.0804%) [255]. This demonstrates that the theoretical precision of $^{13}C$ abundance extracted via partial least squares regression can reach approximately 10‰ even with single-shot emission measurements, which can be sufficient to distinguish variations in the $^{13}C$ isotopes in many naturally occurring carbonaceous materials [256]. In uranium isotope analysis, they also demonstrated that incorporating the UO molecular band at 593.55 nm can improve the precision of $^{235}U$ abundance estimation (expressed as relative standard deviation (RSD)) from 0.72% to 0.42%, compared to combinations of just multiple atomic lines [257].

Surmick *et al* also addressed the non-uniformity in the spatial distribution of plasma temperature by applying spatially resolved molecular spectroscopy (laser: 1064 nm, 8.5 ns, 38 mJ; target: Al alloy 6061; atmosphere: air) [258] using Abel inversion. Harilal *et al* demonstrated the breakdown plasma kernel to be in thermal equilibrium and the off-kernel to be in non-LTE by distinguishing the on- and off-axis molecular emission characteristics (laser: 532 nm, 6 ns, 55 mJ; target: air/Ar/He; atmosphere: vacuum) [259]. Besides that, the enhancement effects of different LIP sources and electron cyclotron resonance microwave discharges on induced plasma generation are also successfully evaluated by measuring vibrational and rotational temperatures from molecular spectra [260-262].

## 4. Numerical advances on different scales in nanosecond laser-target interactions

It is necessary and scientific to model the same physical problem at multiple scales. We take the nanosecond laser-target interaction as an example to understand multi-scale modelling from two aspects. On the one hand, nanosecond laser target interaction involves multiple physical processes and various applications, including pulsed laser deposition, laser-induced breakdown spectroscopy, laser-induced plasma microfabrication, laser shock peening, nanomaterial synthesis, etc. They focus on their own physical and chemical processes and have independent spatial-temporal observation windows.

On the other hand, numerical methods at different scales also have their own inherent simulation intervals accompanied by different temporal and spatial scales [263]. The applicability of the various numerical methods is shown in figure 8. Microscopic methods typified by molecular dynamics simulations can provide detailed insights into individual particles, which are often used to simulate a nanopore or micro-region in the entire workpiece due to computational time and resource constraints [264]. Macroscopic numerical methods for the continuum medium are capable of accurately predicting changes in the density, velocity and pressure fields of fluid flow. Two general discretization approaches are employed in macro-methods. Explicit approach solves unknown variables under the current time step using known variables from the previous time step. The value of the time step cannot be too large due to stability restrictions (such as the Courant-Friedrichs-Lowy (CFL) criterion). In contrast, the implicit approach simultaneously solves a system of discretized equations for all spatial nodes at a given moment in time. To store and solve large matrices (usually sparse matrices) involved is often time-consuming and computationally resource-consuming. Mesoscopic methods based on kinetic theory (grounded in the Boltzmann equation) and statistical physics employ the concept of computational particles (they are often much larger than the actual molecules), whose macroscopic parameters (e.g., velocities and pressures) are obtained by statistical averaging. Meso-scale methods dominate the middle ground between microscopic molecular dynamics and macroscopic methods, which balances computational accuracy and efficiency to a certain extent.

Therefore, the specific strategies of multiple numerical methods are taken to start this section for simulating processes in nanosecond laser-target interactions. At the same time, the great potential of machine learning in numerical simulations has been noted. It will be shown how machine learning can assist in simulating physical processes involved in nanosecond laser-target interactions and the major obstacles that still exist will be discussed, represented by Physics-informed Neural Networks (PINNs).

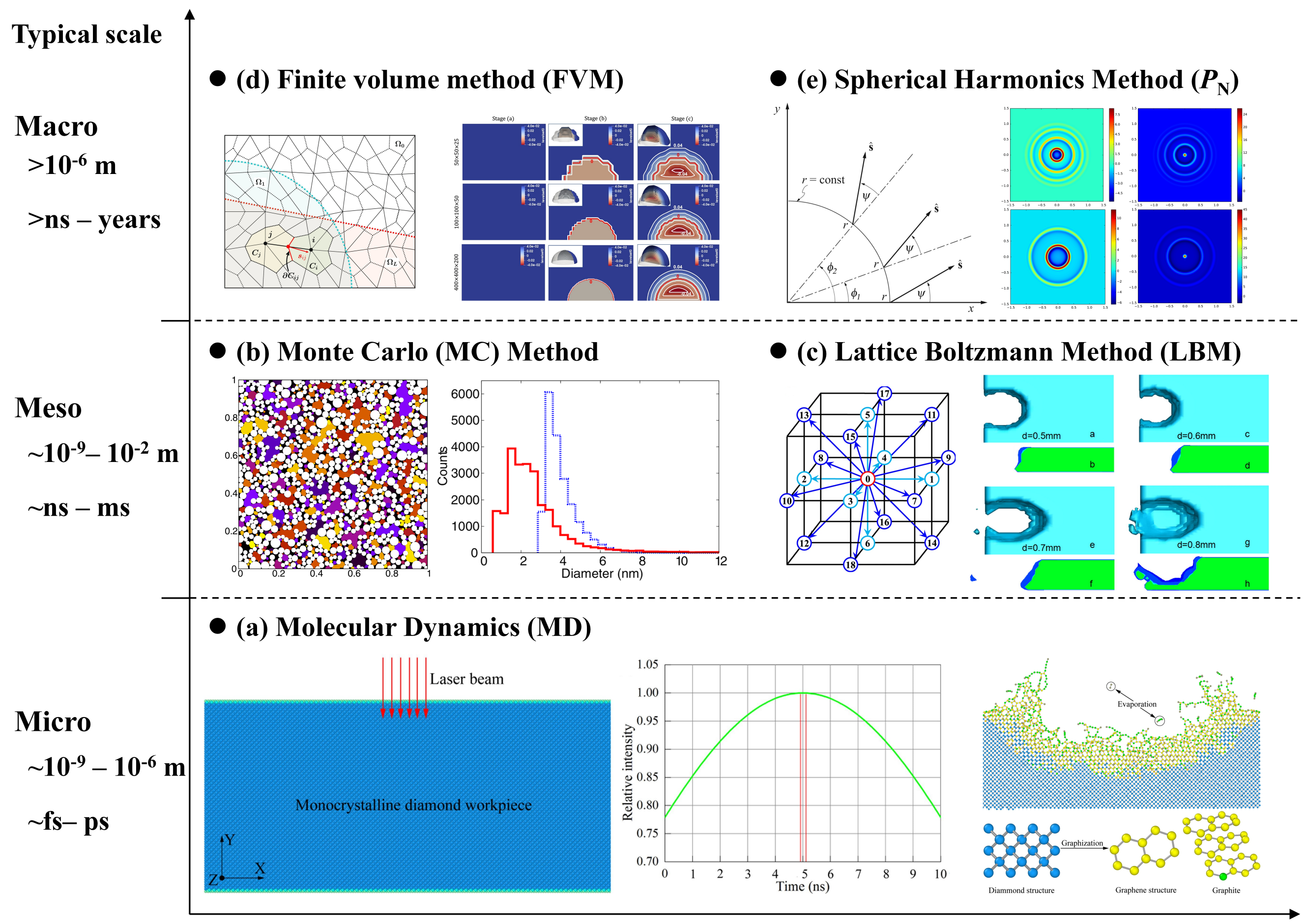

**Figure 8.** Schematic diagram comparing numerical approaches on different scales in nanosecond laser-target interactions. Reproduced from [265]. (a) Microscopic approaches: molecular dynamics: MD model of nanosecond pulsed laser ablation of monocrystalline diamond (left); Gaussian distribution of laser pulse in time domain (median); MD simulation snapshots of laser ablation-induced phase transformation in diamond (right). Reproduced from [266]. (b) Mesoscopic approaches on Monte Carlo method: supercritical bubbles randomly generated during phase explosion in white (left), diameter distribution of bubbles in blue, and droplets in red (right) for Ag, Reproduced from [36]. (c) Lattice Boltzmann method: the D3Q19 discretization scheme for 3D geometry (left), and melt's thickness with different work pieces' thickness (right). Reproduced from [52, 267]. Macroscopic approaches on (d) Finite volume method: 2D illustration of an arbitrary (left) and reinitialization of the level set function after phase transition (right), and (e) Spherical Harmonics Method ($P_N$): Illustration of the invariance of intensity with respect to azimuthal angle $\psi$ at different locations for axisymmetric conditions (left). Example for spherical harmonic approximation of particle density. The degree of approximation is 7 (right). Reproduced from [268-270].

### *4.1 Microscopic approaches based on molecular dynamics*

Molecular dynamics methods have been more widely used to simulate ultrashort pulse laser interactions [271]. Additionally, MD methods provide explicit atomic representations of target heating, vaporization, and plasma plume expansion. Therefore, MD methods are also increasingly being used to explain the nanosecond pulse ablation mechanism at the microscopic scale.

However, since the complete nanosecond pulsed ablation evolves on a much larger spatial and temporal scale than that typically modeled by the conventional MD method, it requires an equivalent conversion in the temporal and spatial domains to be solved first. Zhao *et al* established a three-dimensional MD model for nanosecond laser ablation, predicting the dynamic transition of single-crystal diamond to graphite during ablation by converting a 10 ns pulse duration and 20 μm spot diameter to an equivalent 45 ps pulse duration and 10 nm spot diameter (laser: 355 nm, 10 ns, 50 kHz, 25.0-43.8 J/cm$^2$, 20 μm; target: diamond) [266]. In figure 9, Cao *et al* proposed a multi-scale modelling framework to describe phase explosion ablation and particle/droplet formation in aluminum and copper targets. The model includes two computational domains: smoothed particle hydrodynamics (SPH) and hydrodynamics (HD) (laser: 1064/532 nm, 6/10 ns, 12/24/36 J/cm$^2$; target: Al) [272]. The initial particle distribution for the SPH domain was obtained via MD simulations, with a computational domain size of 20 nm×20 nm×32 μm. Larger atomic distribution mappings were achieved by stitching and combining a range of laser fluxes and beam profiles. Notably, SPH is a mesoscopic particle dynamics method suitable for handling particle sizes ranging from submicron to micron scales, advantageous for dealing with complex geometries without requiring grid partitioning.

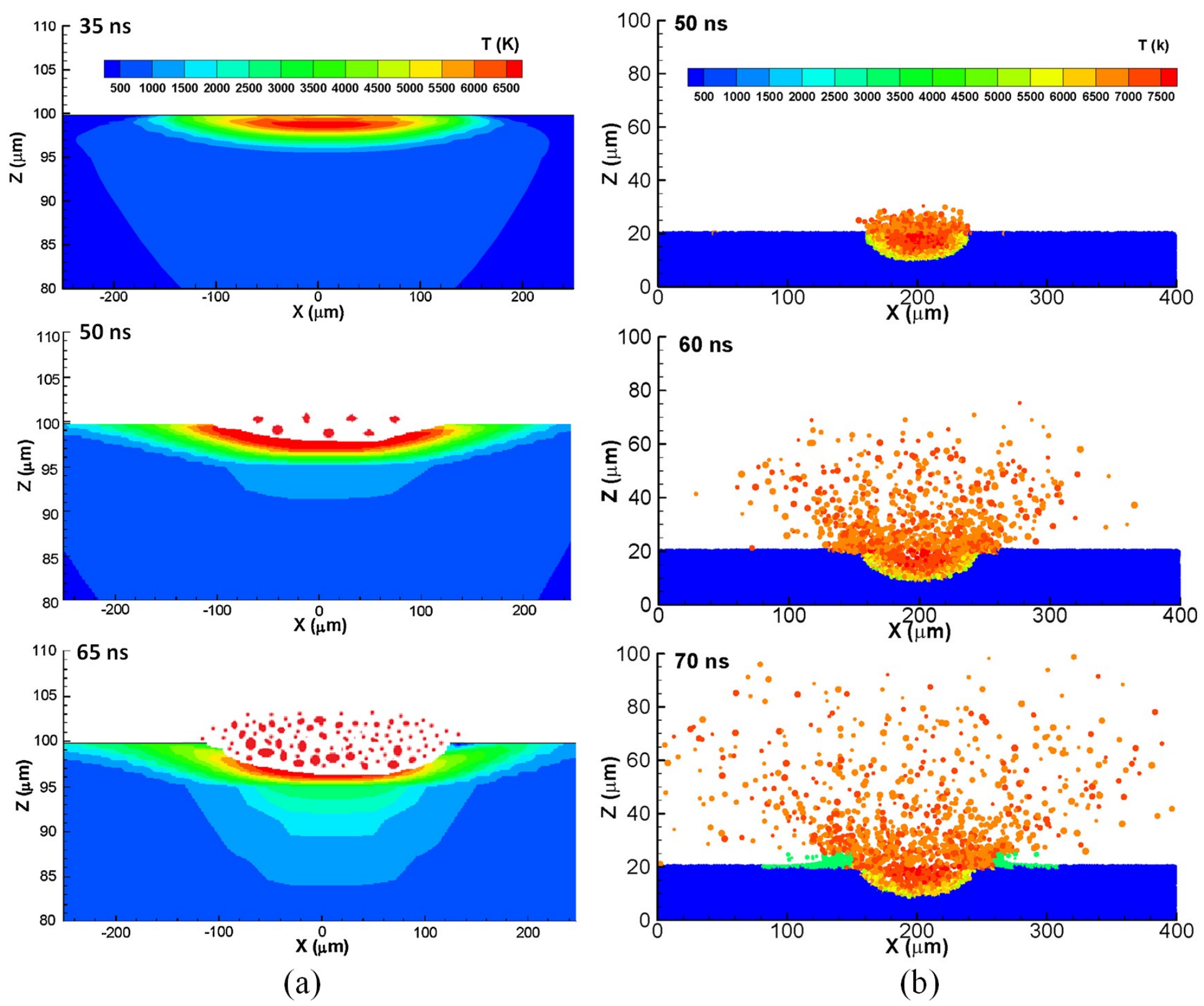

**Figure 9.** The SPH calculation results of the (a) temperature distribution inside the aluminum target, (b) melt ejection for copper at different time (laser fluence 36 J/cm$^2$, wavelength 1064 nm, pulse duration 6 ns, beam diameter 100 μm). (Reproduced from [272]).

### *4.2 Mesoscopic approaches based on kinetic and statistical physics*

#### *4.2.1 Lattice Boltzmann method*

As another popular mesoscopic method, the lattice Boltzmann method is based on the fundamental principle of solving the microscopic kinetic equation of the defined particle distribution functions. One of its advantages lies in its ability to easily simulate multiphase flows involving complex interfaces.

Klassen *et al* used a free surface LBM to simulate the selective laser melting (SLM) process in an Al-Ti binary alloy system. The model used the volume of fluid method to track the free surface between the liquid and gas phases. This model has the ability to predict heat release fluxes on the melting surface, mass loss, and recoil forces generated during vapor expansion [273]. Unlike Free Surface LBM, which separates liquid and gas phases by monitoring mass exchange between adjacent lattices, Pseudopotential LBM introduces a pseudopotential (often referred to as the effective mass) to represent microscopic molecular interactions, allowing automatic phase separation without needing to explicitly capture or track the phase change interface. A general review on pseudopotential LBM can be found in [274]. Zhao *et al* used pseudopotential LBM in combination with enthalpy-based solid-liquid phase change scheme to develop a numerical model for 3D laser cutting of 304 stainless steel, which simulated the melting of the solid workpiece, the interactions between ambient gases and the molten material [52].

*4.2.2 Monte Carlo method*

The MC method is adaptable over a wide range of time scales (from sub-nanosecond to microseconds) and spatial scales (from hundreds of nanometres to hundreds of micrometres). Therefore, the MC method based on stochastic particles has been widely applied to the numerical simulation of microscopic and mesoscopic scale phenomena, such as laser ablation plasma.

Classical MC simulation typically describes collisions between neutral particles, suitable for moderate laser fluxes and lower ionization degrees of plumes. Morozov *et al* conducted a series of studies using the Hard Sphere (HS) collision model. Firstly, they took the delay time of gold evaporation in gold-silver alloys as a free parameter of their model in MC simulations and investigated the effect of delayed evaporation in the alloys on the expansion of the ablation plume in vacuum (laser: 532 nm, 7 ns; target: Au-Ag alloy; atmosphere: vacuum) [43]. Next by assuming uneven distributions of surface temperature and evaporation flux along the irradiated spot radius, they compared the differences in plume expansion velocity and kinetic energy between flat-top and Gaussian beams [275]. Furthermore, on dynamic simulations of laser evaporation into low-pressure and vacuum background atmospheres, they demonstrated a good consistency between MC methods and direct numerical solutions of the Bhatnagar-Gross-Krook (BGK) kinetic equation [276-278]. Different from the HS collision model, Palya *et al* employed a MC method based on Variable Hard Sphere (VHS) collisions to simulate complex shockwave structures in constrained plumes, combining with the Maxwell model of diffuse scattering to describe interactions of particles with multiple wall surfaces (laser: 266 nm, 10 ns, 20 μJ, 20 μm; target: Cu) [187].

On the other hand, since the MC method has the advantage of directly describing particle collisions compared to the continuum methods, Mazzi *et al* combined three-dimensional MC simulations with homogeneous nucleation theory to describe the distribution of liquid nanodroplets and vapor atoms formed during the phase explosion process [279]. The homogeneous nucleation theory was used to evaluate the critical radius for vapor bubble formation as an input parameter required for the MC method, which randomly generates nucleation sites to identify nanodroplets [36].

Moreover, classical MC methods are continuously improved to match the complex evolution behaviors of laser ablation plasma. Ranjbar *et al* utilized a modified MC method incorporating the Saha equation and Lambert-Beer's Law to describe ionization and radiation of plumes and laser absorption, based on the assumption of locally neutral plasma (laser: 266 nm; target: Cu; atmosphere: Ar) [280]. At each time step, the thermal velocity and thermal energy density of the heavy particles are updated by laser absorption, which in turn updates the temperature. This computational model was further utilized to explain phenomena such as the fluid dynamic snowplow effect causing plume splitting [184] and multiple internal shockwaves formation induced by donut beam configurations [281]. Petrov *et al* and Volkov *et al* recognized that conventional MC methods inadequately handle inter-species collisions typically based on semi-empirical Lennard-Jones (L-J) atomic potential parameters using Lorentz-Berthelot mixing rules, proposing the use of Density Functional Theory (DFT) calculations to correct Potential Energy Curve (PEC) for van der Waals (vdW) interactions between atomic pairs [282, 283]. Their calculations indicated the necessity of Morse long-range (MLR) potentials rather than the L-J potential in kinetic simulations involving vapors of substances with non-van der Waals bonding (e.g., metals and semiconductors).

*4.3 Macroscopic approaches based on fluid dynamics*

Among the various macroscopic discrete methods, finite difference (FDM), finite volume (FVM) and finite element (FEM) methods have been widely used in numerical calculations related to laser ablation plasmas, which involve processes such as ablation, mechanical transport and hydrodynamics.

Due to its straightforward discretization, FDM achieves efficient computation in one-dimensional or two-dimensional regular geometric problems. Jain *et al* used an explicit finite difference method to solve the one-dimensional time-dependent heat flow equation considering the latent heat of melting, solidification, and evaporation to simulate the process of pulsed laser irradiation of aluminum and a chromium film being deposited on the aluminum. They assumed that the surface temperature of the melt would not exceed the boiling point of the material [284]. Sawyer *et al* established a laser ablation FDM considering molten flow and coupled mechanical recoil pressure, showing good agreement between simulation and experimental results [285]. Their model employs central differencing for discretization of the heat diffusion and the diffusion-convection terms in the Navier-Stokes equations. Min *et al* (laser: 1064 nm, 10 ns, 10 Hz. 0.5 GW/cm$^2$; target: Al; atmosphere: air ($5\times10^{-4}$-$10^5$ Pa)) [195] used an Alternating Direction Implicit (ADI) scheme in their two-dimensional radiative fluid model to solve the heat conduction equation, and solved the fluid equations using a Semi-Discrete scheme and Runge-Kutta method. The model utilized the Van Leer Flux-Vector Splitting operator splitting algorithm to reduce numerical oscillations encountered during simulations, by separating the computation of conservation quantities from the material transport.

A primary issue within FDM is that mass conservation is only guaranteed as the grid spacing tends to zero, and it struggles with irregular geometries. In contrast, FVM and FEM can handle irregular meshes and relatively complex geometries more efficiently. Gao *et al* analyzed transient thermal-structural coupling of laser-ablated aluminum alloy using FEM with ABAQUS (laser: 1064 nm, 6 ns, 2 J/cm$^2$, 480 µm; target: AA6061) [286]. Luo *et al* analyzed the temperature and stress fields of three-dimensional nonlinear laser-ablated Sn-Te powder based on FEM (laser: 1064 nm, 100 µm; target: SnTe; atmosphere: Ar) [287]. They employed the birth and death of element method in computations to neglect Sn-Te powder in stress fields. FVM emphasizes conservation equations and flux computation, making it more suitable for simulating macroscopic multiphase flows and flow transport processes. Zhao *et al* used the FVM and the exact two-phase Riemann solver (FIVER) to solve the Euler equations. They studied the coupled control equations of laser and fluid by constructing and solving a one-dimensional dual material Riemann problem [268]. In their specific research, they embedded the laser radiation domain boundaries into the fluid domain using the immersed boundary method to simulate the fluid domain affected by the laser radiation. And further to track gas-liquid interfaces, they solved the level set equation with the third-order upwind finite difference format. Finko *et al* simulated two-dimensional inviscid compressible fluid uranium plasma using FVM in Ansys-Fluent. The Pressure-Implicit with Splitting of Operators (PISO) scheme segregated the pressure-based solver from the pressure-velocity coupling [240].

In fluid dynamics computations, strong shockwaves often lead to discontinuities and non-continuity in physical quantities, causing significant “oscillations” and numerical diffusion. The two prevailing methods to handle these problems are shock-fitting and shock-capturing, which also briefly mentioned in Section 2.3.3. Since shock-fitting is more compatible for simulating high Reynolds number flows [288], shock-capturing technique is commonly used for hydrodynamic calculations of LIP more often accompanied by low Reynolds flows. The key to shock-capturing technique is the numerical dissipation. One solution of adding artificial viscosity dissipation terms in fluid equations was originally proposed by von Neumann and Richtmyer in the 1950s. Thereafter, Huerta *et al* introduced a high-order discontinuous Galerkin method for shock-capturing technology [289]. It automatically modifies continuous high-order shape functions to discontinuous ones using a discontinuity sensor based on Mach numbers as sensing variables. Guo *et al* developed a hybrid shock-capturing format based on the WCNS-CPR format, incorporating nonlinear mechanisms such as shock detection and shock capture, demonstrating good performance in accuracy, efficiency, and shock-capturing capability [290]. This technology uses high-order FEM in smooth regions while utilizing strong shock-capturing capabilities provided by FDM or FVM.

To ensure closure of the fluid dynamics equations, an equation of state is indispensable. The SESAME EOS library from Los Alamos National Laboratory or the LEOS library from Lawrence Livermore National Laboratory provides various EOS for about 150 materials, including single elements, mixtures, and compounds. Additionally, Heltem *et al* developed the BADGER EOS library in Fortran, utilizing scaling binding energy for ionic EOS and t-split screened hydrogen model for electronic EOS [291]. The ideal gas model can also be used to solve the BADGER state equation for two particles. Faik *et al* developed the Frankfurt EOS library (FEOS) [292] based on the QEOS model proposed by More *et al* [293] deriving thermodynamic quantities such as pressure, specific internal energy, and enthalpy of arbitrary chemical elements or uniformly mixed elements from specific Helmholtz free energy. It can be used to calculate thermodynamic quantities as a function of density and temperature for arbitrary chemical elements or homogeneous mixtures of elements, and can be programmed into the solution of the fluid equations using an interpolation scheme. Shabanov *et al* proposed a novel method for calculating EOS containing atomic, molecular, and positive and negative ions, employing different solution methods for various system temperatures across a range from 100,000 Kelvin to room temperature [236].

Radiation transport equations are typically solved using the Discrete Ordinates Method ($S_N$) and Spherical Harmonics Method ($P_N$). The primary drawback of the Discrete Ordinates Method is ray effects. Ray effects occur when there is insufficient angular flux resolution in discrete coordinate methods, leading to non-physical artefacts that may cause substantial errors in local solutions. Christensen *et al* proposed the Surface Method mitigating ray effects based on the Discrete Ordinates Method ($S_N$) to solve non-collision fluxes [294]. It tracks a set of ray points on each region surface to compute currents and subsequently obtain flux. Frank *et al* introduced the method of artificial scattering to eliminate ray effects in $S_N$, achieving the same accuracy as $S_N$ but with fewer angular coordinates [295]. Specifically speaking, this method introduces an artificial forward peak scattering operator in radiative transport equations, similar to artificial viscosity in spatial discretization.

As another numerical method for solving for radiation transport, the $P_N$ uses spherical harmonics as spectral functions in angular space. Although $P_N$ is unaffected by ray effects, its accuracy is limited by the substantial computational cost of higher-order expansions. Dai *et al* employed a combined $P_N$ and $S_N$ approach to solve radiative transport equations based on the First Collision Source (FCS) method to suppress ray effects [296]. This method balances computational efficiency and accuracy, using low-order $P_N$ angular expansions to achieve balanced FCS distributions and sufficient angular resolution for collision fluxes. Zhang *et al* overcame inaccuracies of the traditional Spherical Harmonics Discrete Ordinates Method (SHDOM) in describing particle multi-scatter source functions using statistical average scattering phase functions of particle clusters [297].

### *4.4 Advancing numerical simulations with machine learning*

In recent years, with the development of artificial intelligence, machine learning has garnered widespread attention in numerical computational and hydrodynamics communities [298-300], owing to its remarkable data-driven predictive capabilities and strong potential for solving various types of partial differential equations (PDEs). In addressing the forward problems of solving PDEs, machine learning/artificial intelligence has been found capable of accurately solving high-dimensional free boundary PDEs up to 200 dimensions [301], accelerating computation speed and reducing computational time [1], inferring velocity and pressure field distributions from passive scalar concentration fields [302], providing extrapolation in temporal domains [303], and exhibiting robust performance with respect to initial or boundary conditions [304, 305]. In tackling inverse problems such as discovering model parameters within PDEs, machine learning/artificial intelligence has also been indicated to faithfully reconstruct noisy data (including denoising and high-resolution predictions) [306], further predicting model parameters [307] and even deducing closed-form solutions of the defined governing equations [308].

A milestone in this regard is the concept of Physics-informed Neural Networks introduced by the Karniadakis group at the Applied Mathematics Department at Brown University [309], and the base framework is shown in figure 10. This approach utilizes automatic differentiation to express differential operators, integrating physical constraints (e.g., fluid dynamic equations, thermodynamic equations, etc.) into the training of neural networks to solve forward or inverse problems of PDEs [310]. It has been considered to inherit both the interpretability of physical laws and the efficient integration capability with massive data. PINNs have been successfully applied in various domains such as fluid dynamics [311-313], heat transfer [314-316], solid mechanics [317-319], and multi-physics fields coupling modeling [320]. Specifically, within the PINNs framework, physical constraints are incorporated into the loss function of neural networks, typically involving regression residuals of PDEs at specific point sets and residuals of initial and boundary conditions. More comprehensive introductions to the versatility and latest advancements of PINNs can be found in [321-323]. Some solvers based on the PINNs framework are summarized in table 2.

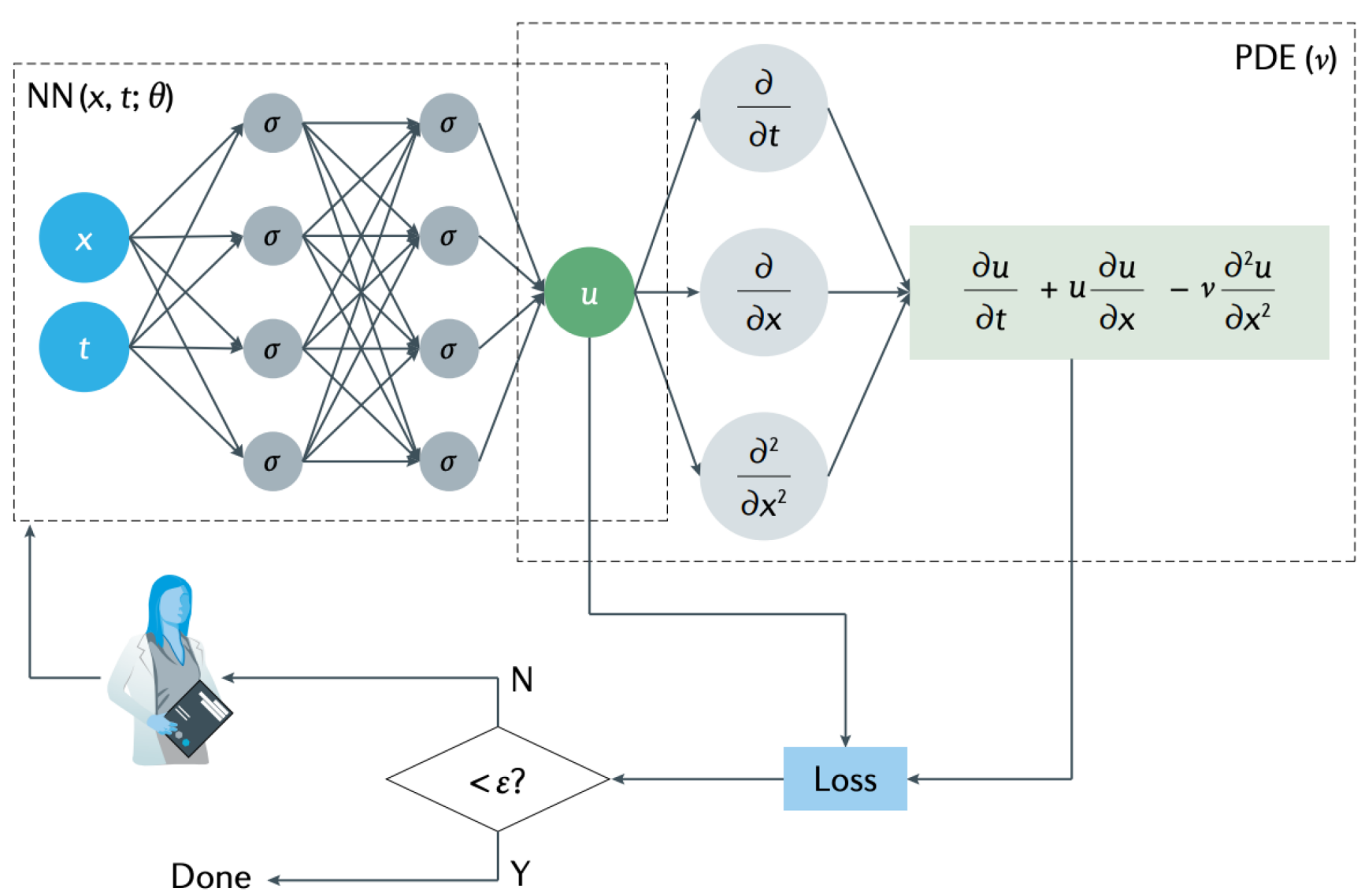

**Figure 10.** Schematic of PINN algorithm. Reprinted from [310]. Copyright 2021 with permission from Springer Nature Limited.

**Table 2.** PINNs frame works being active developed. A solver solves the problem defined by users. A wrapper does not solve, but only wraps low-level functions from other libraries (e.g., PyTorch) into high-level functions that are convenient for users to implement PINN to solve the problem. Reproduced from [298].

| Framework site | Usage | Language | Backend |
|---|---|---|---|
| DeepXDE | Solver | Python | TensorFlow, PyTorch, JAX, Paddle |
| NVIDA Modulus (SimNet) | Solver | Python | TensorFlow |
| PyDEns | Solver | Python | TensorFlow |
| NeuroDiffEq | Solver | Python | PyTorch |
| NeuralPDE | Solver | Julia | Julia |
| SciANN | Wrapper | Python | TensorFlow |
| ADCME | Wrapper | Julia | TensorFlow |

| GPytorch | Wrapper | Python | PyTorch |
| --- | --- | --- | --- |
| Neural Tangents | Wrapper | Python | JAX |

In recent years, PINNs have begun to support research goals in numerical computations of laser ablation and plasma physics. **In thermal simulations of laser-material interactions**, by embedding the heat conduction equation into the loss function of neural networks, Voigt *et al* employed PINNs to solve dynamic heat conduction in block materials under Gaussian profile beam irradiation, achieving results consistent with FEM outcomes [324]. Chen *et al* used PINNs to predict time-varying temperature distributions of metal powders during Laser Powder Bed Fusion (L-PBF) additive manufacturing processes [325]. Xie *et al* demonstrated that PINNs achieve the same accuracy with minimal training data in predicting 3D temperature field distributions during Direct Energy Deposition (DED) laser additive manufacturing processes, compared to other data-driven machine learning methods such as DNNs, LSTMs, and XGBoost, attributed to the advantage of PINNs in physical constraints. Additionally, Bowman *et al* compared prediction errors and convergence rates of different types of activation functions and their combinations in predicting laser-skin tissue interactions under adiabatic and convective boundaries [326]. **In simulating thermo-mechanical coupling of lasers with materials**, by simultaneously embedding the dimensionless heat conduction and thermal stress equations into the loss function of neural networks, You *et al* combined transfer learning to enable PINNs to more efficiently predict thermal displacements, thermal stresses and thermal strains on Thin Disk Laser Crystals (TDLCs) with varying thermal conductivities and thicknesses [327]. **In plasma simulation**, Zhong *et al* proposed two PINN-based frameworks for low-temperature plasma simulations to solve plasma equations with solution-dependent coefficients and transient terms [328]. Besides this, works related to plasma simulation have focused on discharge and arc plasmas [329, 330], fusion plasmas [331, 332], magnetohydrodynamics [333-335], and space and astrophysical plasmas [301-306].

On the other hand, less attention is focused on machine learning-assisted plasma radiation calculations. Among these studies, Kluth *et al* explored a deep learning-accelerated computational framework for collisional-radiative model calculations [336], while Datta *et al* compared various machine learning algorithms effectiveness in predicting electron temperature and ion number densities using spectral data generated by collisional-radiative codes [337]. Ghosh *et al* focused on using deep neural networks to learn molecular spectra based on input molecular structures [338]. Similarly, there is a call for more attention in plasma spectroscopy calculations, with a potential framework being the integration of PINN concepts, where the radiation transport equations describing atomic or molecular spectra could be part of the loss function, aiding in solving radiation transport equations with neural network assistance to avoid complex grid divisions in numerical computations.

Finally, some possible reasons why machine learning appears to struggle when assisting in the numerical modeling of nanosecond laser ablation plasmas are listed: (i) In terms of spatial domains, the physical modeling of nanosecond laser ablation plasmas inherently involves the coupled computation of multiple spatial domains and they are dominated by different controlling equations. At the same time, (ii) in terms of temporal scales, the relevant physical quantities evolve drastically, such mutant data in temporal and spatial domains can make the inherent Vanishing and Exploding Gradients Problems in neural networks even more intractable. And (iii) in terms of physical effects, the bidirectional coupling of multiple physicochemical effects cannot be known beforehand which ones can be neglected, e.g. thermal, mechanical, fluid, radiation effects. (iv) In terms of fundamental principles, just as the theoretical advances in nanosecond laser ablation plasma introduced in the previous section, different models have their own applicable explanation intervals along with their respective assumptions and inadequacies. Therefore, there is still a long way to go for researchers to get closer to the “truth” behind the nanosecond laser-target interactions.

These factors make the construction of neural network frameworks need to be designed and handled exceptionally carefully (e.g., activation function (which is one of the factors preventing data-driven machine learning algorithms from being more physically relevant); number, thresholds and weights of artificial neurons or nodes; and also, iterative solution algorithms). More interdisciplinary researchers need to be called upon to focus on the artificial intelligence modelling of the nanosecond laser ablation plasma. To conclude, some examples of improved numerical approaches in laser processing, nanoparticle preparation, fundamental research in laser-induced plasma, and so on are summarized in table 3 as the end.

**Table 3.** Summary of recent algorithm improvements in macro/micro/mesoscale simulation of laser processing, nanoparticle preparation, laser-induced plasma, etc.

| Micro/Meso/Macro-level | | Improved approach | Application | Reference |
|---|---|---|---|---|
| **Micro-level** | MD | The velocity reconstruction algorithm is used to change the time scale and predict the material phase transition process with smaller time scale. | Precision machining | [266] |
| | | MD and finite difference methods are used to calculate phase explosion and particle formation respectively to form a multi-scale model. | Nanomaterials preparation | [272] |
| **Meso-level** | SPH | The SPH method within the Arbitrary Lagrangian-Eulerian (ALE) framework is utilized in order to model the waves travelling inside the liquid metal droplet impacted by a laser pulse and to identify the two cavitation regimes. | Laser processing | [339] |
| | LBM | The gas-liquid interface of Al-Ti binary alloy is traced by free surface flow method. | Laser ablation | [273] |
| | | The pseudo-potential (effective mass) is introduced to represent the microscopic molecular interaction. | Multiphase flow | [274] |
| | MC | The evaporation delay time of gold and silver is added to the evaporation criterion of gold and silver alloy to simulate the fractionation caused by melting point difference. | Laser processing | [43] |
| | | It is assumed that the surface temperature and evaporation flux are non-uniformly distributed over the irradiation-spot surface for a Gaussian beam. | Laser ablation | [275] |
| | | The Maxwell model of diffuse scattering describes the interaction between particles and walls. | Laser induced plume evolution | [187] |
| | | Homogeneous nucleation theory is used as a criterion for the formation of phase explosion nanoparticles. | Nanoparticle preparation | [279] |
| | | Density functional theory (DFT) corrects the van der Waals interaction of two-body pairs of atoms. | Laser processing | [282, 283] |
| **Macro-level** | FDM | The central difference scheme deals with convection terms in thermal diffusion and fluid expansion. | Laser ablation | [285] |
| | | The interaction of two carbon fluids is simulated by electrostatic force and collision term interaction. | Laser-induced plasma | [340] |
| | FEM | The grid birth and death of element method can ignore the calculation of the stress field of the material powder. | Precision machining | [287] |
| | | The electric field under the defect is solved by Helmholtz equation, and the laser beam distribution under the defect is described by the square relation. | Laser processing | [341] |
| | | JmatPro software is used to calculate the temperature dependence of the thermal and mechanical properties of the material, and the material parameters were temperature dependent. | Laser processing | [286] |
| | FVM | The boundary of the laser radiation domain is embedded into the fluid domain by the embedding boundary method, and the gas-liquid interface is tracked by the level set method. | Laser induced cavitation bubbles | [268] |
| | | The pressure implicit splitting operator is used to separate the pressure-velocity coupling. | Laser induced plume evolution | [240] |
| **AI advancing** | PINN | Damage function of heat conduction equation embedded in neural network. | Additive manufacturing | [324-326] |
| | | Dimensionless heat conduction and thermal stress equations are simultaneously embedded into the damage function of the neural network. | Laser processing | [327] |
| | | The neural network solves the plasma governing equation. | Laser-induced plasma | [328] |
| | Deep learning | Accelerated collision radiation model calculation framework. | Collisional-radiative model | [336] |
| | | Deep learning establishes associations between molecular structures and molecular spectra. | Molecular spectra | [338] |
| | Machine learning | Machine learning algorithms use radiation spectra to predict electron temperature and ion number density. | LIBS | [337] |

## 5. Conclusions and perspectives

It is always crucial to carefully account for different physical processes to understand and optimize experimental parameters for industrial applications, and with this goal in mind, recent progress in the numerical modeling and simulation on the nanosecond laser and the solid target interactions are presented in this review.

The multiphase real EOS and temperature-dependent material properties (including optical absorption coefficient, density, thermal conductivity, specific heat, and the like) are gradually applied to improve the ablation process calculation. Multi-species with non-stoichiometry are considered in the mass and energy transfer from target to plasma, which directly determine the species as well as the velocity distribution within the initial plasma. The mechanistic coupling effects of target-plasma-atmosphere parameters are gaining experimental and industrial applications while the search for the physical reasons behind are also being reawakened. At the same time, the influence of various transport phenomena and the EOS of real gases are used to improve the modeling of plasma expansion, while the effects of plasma radiation and external magnetic field on plasma expansion have also gained more attention and are further considered quantitatively. The exploration of both equilibrium/non-equilibrium atomic processes and chemical reactions in LIP is also advancing, with researchers from quantum mechanics and theoretical physics providing solutions to more accurate ionization/collision cross sections and rate coefficients, both from theoretical and experimental perspectives. In particular, four powerful modeling approaches are discussed separately for explaining ion acceleration processes within nanosecond LIP at moderate laser energy inputs. The exploration of atomic and molecular emission for elemental analysis and evaluation of fundamental physical processes are also emphasized. In addition, specific numerical modeling approaches on different scales are compiled, while combining machine learning to optimize and accelerate numerical calculations provides an interesting direction to be further explored with its advantages of efficient data integration.

However, the numerical modeling and simulation on the nanosecond laser and the solid target interactions remain challenging. First, the physical understanding and numerical modeling of nanosecond-laser interactions are still a greater challenge because it involves multi-physics, multi-space and multi-object coupled processes with parameters that change rapidly in time and space. The application of the wide-range multiphase EOS and material parameters along with higher accuracy is crucial for modelling non-equilibrium and non-steady-state physical processes in the laser-target interactions at the same level.

Second, some unique physical phenomena and laws behind have received relatively insignificant attention and still need to be investigated as important research subjects. For example, there is evidence of a link between laser ablation characteristics and the thermal and mechanical properties of materials. The external magnetic field has a significant contribution to the expansion and ionization evolution of the plasma. Ion acceleration is widespread in multicomponent plasma plumes.

Thirdly, although numerical modeling approaches at different scales are widely discussed and adopted, the connection and combination between them remains to be continuously investigated. There is no doubt that almost all problems have multiple scales in space and time, and a single scale modeling is not sufficient to completely address multiscale problems, where multiscale simulations provide better understanding under a unified framework. At the same time, the training and building of machine learning frameworks needs more efforts to be invested in order to provide more practical experiences, as this shows a great potential in modeling nanosecond laser-target interactions from two aspects: i) solving the physics equations efficiently, ii) backpropagating the model parameters from the experimental/computational results.

## Acknowledgement

This work was supported in part by the National Key Research and Development Plan of China (No. 2021YFB3703200), and in part by CNNC Science Fund for Talented Young Scholars.

## Data availability statement

All data that support the findings of this study are included within the article (and any supplementary files).